\documentclass[twocolumn,showpacs,preprintnumbers,amsmath,amssymb]{revtex4-1}

\usepackage{amsfonts}
\usepackage{amsmath}
\usepackage{amssymb}
\usepackage{graphicx}
\usepackage{dcolumn}
\usepackage{bm}

\begin{document}


\title{Reciprocity of weighted networks}

\author{Tiziano Squartini}
\affiliation{Instituut-Lorentz for Theoretical Physics, Leiden Institute of Physics, University of Leiden, Niels Bohrweg 2, 2333 CA Leiden, The Netherlands }
\author{Francesco Picciolo}
\affiliation{Department of Chemistry, University of Siena, Via A. De Gasperi 2, 53100 Siena, Italy }
\author{Franco Ruzzenenti}
\affiliation{Department of Chemistry, University of Siena, Via A. De Gasperi 2, 53100 Siena, Italy}
\author{Diego Garlaschelli}
\affiliation{Instituut-Lorentz for Theoretical Physics, Leiden Institute of Physics, University of Leiden, Niels Bohrweg 2, 2333 CA Leiden, The Netherlands}


\date{\today}

\begin{abstract}
In directed networks, reciprocal links have dramatic effects on dynamical processes, network growth, and higher-order structures such as motifs and communities. While the reciprocity of binary networks has been extensively studied, that of weighted networks is still poorly understood, implying an ever-increasing gap between the availability of weighted network data and our understanding of their dyadic properties. Here we introduce a general approach to the reciprocity of weighted networks, and define quantities and null models that consistently capture empirical reciprocity patterns at different structural levels. We show that, counter-intuitively, previous reciprocity measures based on the similarity of mutual weights are uninformative. By contrast, our measures allow to consistently classify different weighted networks according to their reciprocity, track the evolution of a network's reciprocity over time, identify patterns at the level of dyads and vertices, and distinguish the effects of flux (im)balances or other (a)symmetries from a true tendency towards (anti-)reciprocation.
\end{abstract}

\pacs{Valid PACS appear here}

\maketitle

\section*{Introduction}

The study of \emph{link reciprocity} in binary directed networks \cite{HL,WF}, or the tendency of vertex pairs to form mutual connections, has received an increasing attention in recent years \cite{myreciprocity,mygrandcanonical,newmanpropagation,
serranopercolation,vinkolocalization,vinkowiki,myalessandria,mysymmetry2,vinko1,vinko2,foodwebmotifs,mymethod}. Among other things, reciprocity has been shown to be crucial in order to classify \cite{myreciprocity} and model \cite{mygrandcanonical} directed networks, understand the effects of network structure on dynamical processes (e.g. diffusion or percolation processes \cite{newmanpropagation,serranopercolation,vinkolocalization}), explain patterns of growth in out-of-equilibrium networks (as in the case of the Wikipedia \cite{vinkowiki} or the World Trade Web \cite{myalessandria,mysymmetry2}), and study the onset of higher-order structures such as correlations \cite{vinko1,vinko2} and triadic motifs  \cite{motifs,foodwebmotifs,mymethod,mytriadic}. 
In networks that aggregate temporal information such as e-mail or phone-call networks, reciprocity also provides a measure of the simplest \emph{feed-back} process occurring in the network, i.e. the tendency of a vertex to \emph{respond} to another vertex stimulus. 
Finally, reciprocity quantifies the information loss determined by projecting a directed network into an undirected one: if the reciprocity of the original network is maximum, the full directed information can be retrieved from the undirected projection; on the other hand, no reciprocity implies a maximum uncertainty about the directionality of the original links that have been converted into undirected ones \cite{myreciprocity}. In particular intermediate cases, significant directed information can be retrieved from an undirected projection using the knowledge of reciprocity \cite{myalessandria}. 
In general, reciprocity is the main quantity characterizing the possible dyadic patterns, i.e. the possible types of connections between two vertices. 

While the reciprocity of binary networks has been studied extensively, that of weighted networks has received much less attention \cite{jari,fagiolo,achen,faloutsos}, because of a more complicated phenomenology at the dyadic level.
While in a binary graph it is straightforward to say that a link from vertex $i$ to vertex $j$ is reciprocated if the link from $j$ to $i$ is also there, in a weighted network there are clear complications.
Given a link of weight $w_{ij}>0$ from vertex $i$ to vertex $j$, how can we assess, in terms of the mutual link of weight $w_{ji}$, whether the interaction is reciprocated?
While $w_{ji}=0$ (no link from $j$ to $i$) clearly signals the absence of reciprocation, what about a value $w_{ji}>0$ but such that $w_{ji}\ll w_{ij}$?
This complication has generally led to two approaches to the study of directionality in weighted networks: one assuming (either explicitly or implicitly) that perfect reciprocity corresponds to symmetric weights ($w_{ij}= w_{ji}$) \cite{jari,achen,faloutsos}, and one looking for deviations from such symmetry by studying net flows (or imbalances), defined as $w_{ij}-w_{ji}$ \cite{serranoimbalance}. 
In the latter approach, significant information about the original weights, including their reciprocity, is lost: the original network produces the same results as any other network where $w'_{ij}=w_{ij}+\Delta_{ij}$ and $w'_{ji}=w_{ji}+\Delta_{ij}$. Since $\Delta_{ij}$ is arbitrary, this approach cannot distinguish networks that have very different symmetry properties. In particular, maximally asymmetric (i.e. $\Delta_{ij}=-w_{ji}$, implying $w'_{ji}=0$ whenever $w'_{ij}>0$) and maximally symmetric networks (i.e. $\Delta_{ij}\gg w_{ij}+w_{ji}$, implying $w'_{ij}\approx w'_{ji}$), which are treated as opposite in the first approach, are indistinguishable in the second one. Consider, for example, two nodes $a$ and $b$ linked by the asymmetric weights $w_{ab}=0$ and $w_{ba}=10$: the imbalance $w_{ba}-w_{ab}$ is the same as if they were an almost symmetric dyad with $w_{ab}=10^4$ and $w_{ba}=10^4+10\simeq 10^4$.

In addition to the above limitations, it has become increasingly clear that the heterogeneity of vertices, which in weighted networks is primarily reflected into a generally very broad distribution of the \emph{strength} (total weight of the links entering or exiting a vertex \cite{vespy_weighted}), must be taken into account in order to build an adequate null model of a network \cite{MS,MSZ,generating,chung_lu,newman_origin,katanza,
newman_expo,weightedconfiguration,serrano_weighted,
serrano_richclub,colizza_richclub,mybosefermi}.
Indeed, the different intrinsic tendencies of individual vertices to establish and/or strengthen connections have a strong impact on many other structural properties, and the reciprocity is no exception. 
It is therefore important to account for such irreducible heterogeneity by treating local properties such as the strength (or the degree in the binary case) as constraints defining a null model for the network \cite{mymethod}. 
While null models of weighted networks are generally computationally demanding \cite{serrano_richclub,colizza_richclub}, recently a fast and analytical method providing exact expressions characterizing both binary and weighted networks with constraints has been proposed \cite{mymethod}.
This allows us, for the first time, to have mathematical expressions characterizing the behaviour of topological properties under the null model considered.
In this paper we extend those results, in order to propose new mathematical definitions of reciprocity in the weighted case and to evaluate their behaviour exactly under various null models that introduce different constraints.
This also allows us to assess whether an observed asymmetry between reciprocal links is consistent with fluctuations around a balanced but noisy average, or whether it a statistically robust signature of imbalance.
Finally, we introduce models that successfully reproduce the observed patterns by introducing either a correct global reciprocity level or more stringent constraints on the local reciprocity structure.

\section*{Results\label{sec_res}}

We first introduce measures of reciprocity which meet three criteria simultaneously: 1) if applied to a binary network, they must reduce to their well-known unweighted counterparts; 2) they must allow a consistent analysis across all structural levels, from dyad-specific through vertex-specific to network-wide; 3) they must have a mathematically controlled behaviour under null models with different constraints, thus disentangling reciprocity from other sources of (a)symmetry. 
Then, we discuss the differences with respect to other inadequate measures of `symmetry', show our empirical results, and introduce theoretical models aimed at reproducing the reciprocity structure of real weighted networks.

\subsection*{Dyad-specific measures}

\indent We consider a directed weighted network specified by the weight matrix $W$, where the entry $w_{ij}$ indicates the weight of the directed link from vertex $i$ to vertex $j$, including the case $w_{ij}=0$ indicating the absence of such link. For simplicity, we assume no self-loops (i.e. $w_{ii}=0$ $\forall i$), as the latter carry no information about reciprocity (in any case, allowing for self-loops is straightforward in our approach).
As Fig. 1 shows, we can always decompose each pair $(w_{ij},w_{ji})$ of reciprocal links into a bidirectional (fully reciprocated) interaction, plus a unidirectional (non reciprocated) interaction.

\begin{figure}
\includegraphics[width=.47\textwidth]{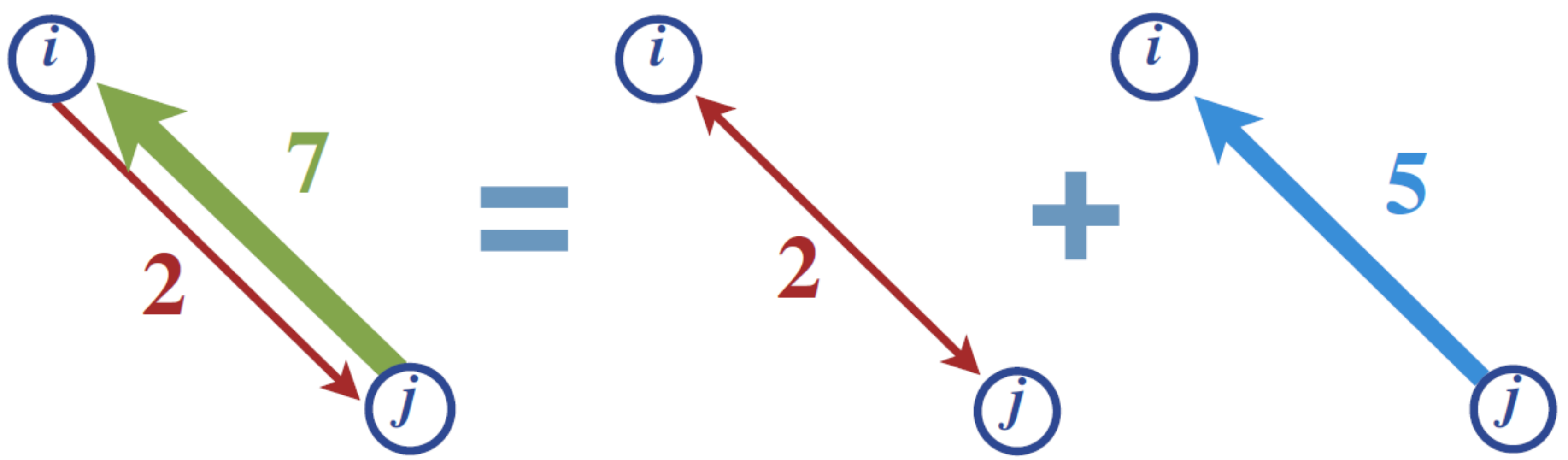}
\caption{Basic decomposition of any two dyadic fluxes (in the example shown, $w_{ij}=2$ and $w_{ji}=7$) into a fully reciprocated component ($w_{ij}^\leftrightarrow=2$) and a fully non-reciprocated component ($w_{ij}^\leftarrow=5$, which implies $w_{ij}^\rightarrow=0$).}
\end{figure}

Formally, we can define the \emph{reciprocated} weight between $i$ and $j$ (the symmetric part) as
\begin{equation}
w^\leftrightarrow_{ij}\equiv\min[w_{ij},w_{ji}]=w^\leftrightarrow_{ji}
\label{eq:wlr}
\end{equation}
and the \emph{non-reciprocated weight} from $i$ to $j$ (the asymmetric part) as
\begin{equation}
w^\rightarrow_{ij}\equiv w_{ij}-w^\leftrightarrow_{ij}
\label{eq:wr}
\end{equation}
Note that if $w^\rightarrow_{ij}>0$ then $w^\rightarrow_{ji}=0$, which makes the unidirectionality manifest. 
We can also define 
\begin{equation}
w^\leftarrow_{ij}\equiv w_{ji}-w^\leftrightarrow_{ij}=w^\rightarrow_{ji}
\label{eq:wl}
\end{equation}
as the \emph{non-reciprocated weight} from $j$ to $i$, and restate the unidirectionality property in terms of the fact that $w^\rightarrow_{ij}$ and $w^\leftarrow_{ij}$ cannot be both nonzero.
Thus any dyad  $(w_{ij},w_{ji})$ can be equivalently decomposed as $(w^\leftrightarrow_{ij},w^\rightarrow_{ij},w^\leftarrow_{ij})$.
If the network is binary, all the above variables are either $0$ or $1$ and our decomposition coincides with a well studied dyadic decomposition \cite{myreciprocity,mygrandcanonical,newmanpropagation,
serranopercolation}.

\subsection*{Vertex-specific measures}

\indent From the above fundamental dyadic quantities it is possible to define reciprocity measures at the more aggregate level of vertices. We recall that the out- and in-strength of a vertex $i$ are defined as the sum of the weights of the out-going and in-coming links respectively:
\begin{equation}
s^{out}_i=\sum_{j\ne i}w_{ij}\qquad s^{in}_i=\sum_{j\ne i}w_{ji}
\end{equation}
In analogy with the so-called degree sequence in binary networks, we denote the vector of values $\{s^{out}_i\}$ as the \emph{out-strength sequence}, and the vector of values $\{s^{in}_i\}$ as the \emph{in-strength sequence}.
Using eqs.(\ref{eq:wlr}-\ref{eq:wl}), we can split the above quantities into their reciprocated and non-reciprocated contributions, as has been proposed for vertex degrees in binary networks \cite{myreciprocity,mygrandcanonical}. We first define the \emph{reciprocated strength}
\begin{equation}
s^\leftrightarrow_{i}\equiv\sum_{j\ne i}w^\leftrightarrow_{ij}
\label{eq:slr}
\end{equation}
which measures the overlap between the in-strength and the out-strength of vertex $i$, i.e. the portion of strength of that vertex which is fully reciprocated by its neighbours. Then we define the 
\emph{non-reciprocated out-strength} as
\begin{equation}
s^{\rightarrow}_i\equiv \sum_{j\ne i}w^{\rightarrow}_{ij}=s^{out}_i-s^\leftrightarrow_i
\label{eq:sr}
\end{equation}
and the \emph{non-reciprocated in-strength} as
\begin{equation}
s^{\leftarrow}_i\equiv\sum_{j\ne i}w^{\leftarrow}_{ij}=s^{in}_i-s^\leftrightarrow_i
\label{eq:sl}
\end{equation}
The last two quantities represent the non-reciprocated components of $s^{out}_i$ and $s^{in}_i$ respectively, i.e. the out-going and in-coming fluxes which exceed the inverse fluxes contributed by the neighbours of vertex $i$. 

\subsection*{Network-wide measures}

\indent Finally, we introduce weighted measures of reciprocity at the global, network-wide level. Recall that the total weight of the network is 
\begin{equation}
W\equiv \sum_i\sum_{j\neq i}w_{ij}=\sum_i s^{out}_i=\sum_i s^{in}_i
\end{equation}
Similarly, we denote the \emph{total reciprocated weight} as 
\begin{equation}
W^\leftrightarrow\equiv \sum_i\sum_{j\neq i}w^\leftrightarrow_{ij}=\sum_i s^{\leftrightarrow}_i
\label{eq:wrec}
\end{equation}
Extending a common definition widely used for binary graphs \cite{HL,WF,myreciprocity}, we can then define the \emph{weighted reciprocity} of a weighted network as
\begin{equation}
r\equiv\frac{W^\leftrightarrow}{W}
\label{eq:r}
\end{equation}
If all fluxes are perfectly reciprocated (i.e. $W^\leftrightarrow=W$) then $r=1$, whereas in absence of reciprocation (i.e. $W^\leftrightarrow=0$) then $r=0$.
In the Appendix we discuss the difference between our definitions and other attempts to characterize the reciprocity of weighted networks \cite{jari,fagiolo,achen,faloutsos}. 

Just like its binary counterpart, eq.(\ref{eq:r}) is informative only after a comparison with a null model (NM) is made, i.e. with a value $\langle r\rangle_{NM}$ expected for a network having some property in common (e.g. the number of vertices $N$ and/or the total weight $W$) with the observed one. 
As a consequence, networks with different empirical values of such quantities cannot be consistently ranked in terms of the measured value of $r$. An analogous problem is encountered in the binary case \cite{myreciprocity}, and has been solved by introducing a transformed quantity that we generalize to the present setting as
\begin{equation}
\rho_{NM}\equiv\frac{r-\langle r\rangle_{NM}}{1-\langle r\rangle_{NM}}
\label{eq:rho}
\end{equation}
The sign of $\rho_{NM}$ is directly informative of an increased, with respect to the null model, tendency to reciprocate ($\rho_{NM}>0$) or to avoid reciprocation ($\rho_{NM}<0$). If $\rho_{NM}$ is consistent with zero (within a statistical error that we quantify in the Appendix), then the observed level of reciprocity is compatible with what merely expected by chance under the null model. The literature on null models of networks is very vast \cite{mygrandcanonical,mymethod,MS,MSZ,generating,chung_lu,newman_expo,weightedconfiguration,mybosefermi,bias,myWRG,snijders,pattison,ciccio}. In this paper we adopt a recent analytical method \cite{mymethod} and extend it in order to study the reciprocity of weighted networks.
The three null models we consider are described in the Methods and Appendix.

\subsection*{Reciprocity versus symmetry}

We stress that the alternative approaches are all based on the assumption that the maximum level of reciprocity corresponds to a symmetric network where $w_{ij}=w_{ji}$, so that deviations from this symmetric situation are interpreted as signatures of incomplete reciprocity. 
This is actually incorrect: independently of other properties of the observed network, the symmetry of weights (i.e. $w_{ij}=w_{ji}$) is completely uninformative about the reciprocity structure, for two reasons. 

First, in networks with broadly distributed strengths (as in most real-world cases) the attainable level of symmetry strongly depends on the in- and out-strengths of the end-point vertices: unless $s_i^{in}=s_i^{out}$ for all vertices, it becomes more and more difficult, as the heterogeneity of strengths across vertices increases, to match all the constraints required to ensure that $w_{ij}=w_{ji}$ for all pairs. Therefore, even networks that maximize the level of reciprocity, given the values of the strengths of all vertices, are in general not symmetric. 

On the other hand, in networks with balance of flows at the vertex level ($s_i^{in}=s_i^{out}$ for all vertices) an average symmetry of weights ($\langle w_{ij}\rangle=\langle w_{ji}\rangle$) is automatically achieved by pure chance, even without introducing a tendency to reciprocate (see Appendix). In many real networks (including examples we study below), the balance of flows at the vertex level is actually realized, either exactly or approximately, as the result of conservation laws (e.g. mass or current balance). In those cases, the symmetry of weights should not be interpreted as a preference for reciprocated interactions. 

In the Appendix we also show that measures based on the correlation between $w_{ij}$ and $w_{ji}$ are flawed. Similarly, studies of asymmetry focusing on the differences $w_{ij}-w_{ji}$ are severely limited by the fact that the observed imbalances might actually be fluctuations around a zero average ($\langle w_{ij}-w_{ji}\rangle=0$), irrespective of the level of reciprocity.
Thus, reciprocity and symmetry are two completely different structural aspects \cite{mysymmetry2}. 

\begin{table*}[t!]
\begin{tabular}{lcccc}
\hline
   & $ \rho_{WCM}$ & $\rho_{BCM}$& $\rho_{WRG}$& $r$\\
\hline
\textbf{Social networks} (3 nets) &  &  & & \\
Most reciprocal  & $0.75 \pm 0.01 $ & $0.75 \pm 0.01$& $0.75 \pm 0.01 $& $0.85  \pm 0.01$\\
Least reciprocal  & $0.59 \pm 0.01 $ & $0.58 \pm 0.02$& $0.57 \pm 0.02 $& $ 0.78 \pm 0.01$ \\
\hline
\textbf{World Trade Web} (53 nets) &  & &  & \\
Most reciprocal   & $0.59 \pm 0.03 $ & $0.57 \pm 0.03$& $0.57 \pm 0.04 $ & $0.79 \pm 0.02$\\
Least reciprocal  & $0.43 \pm 0.03 $ & $0.35 \pm 0.05$& $0.36 \pm 0.05 $ & $ 0.66 \pm 0.02$\\
\hline
\textbf{Interbank networks} (5 nets) &  & &  & \\
Most reciprocal  & $0.07 \pm0.02 $ & $-0.26 \pm 0.03$& $-0.26 \pm 0.03 $ & $0.37 \pm 0.01$\\
Least reciprocal  & $0.02 \pm 0.01 $ & $-0.40 \pm 0.02$& $-0.40 \pm 0.02 $ & $ 0.30 \pm 0.01$\\
\hline
\textbf{Neural network} (1 net)& & & &\\
$\mbox{\textit{C. Elegans}}$  & $0.02 \pm 0.01 $ & $-0.11 \pm 0.03$& $-0.007 \pm 0.01$ & $0.08 \pm 0.01$\\
\hline
\textbf{Foodwebs} (8 nets) & &  &  & \\
Most reciprocal  & $-0.14 \pm 0.26 $ & $-0.67 \pm 0.20$& $ -0.65 \pm 0.20 $ & $0.17 \pm 0.02$\\
Least reciprocal & $-0.34 \pm 0.22 $ & $-0.97 \pm 0.02$& $-0.97 \pm 0.02 $  & $ 0.01 \pm 0.02$\\
\hline
\end{tabular}
\caption{Reciprocity of 70 real weighted networks (see the SI for a description of the data), measured using $\rho_{NM}$ under 3 null models (Weighted Configuration Model, Balanced Configuration Model, Weighted Random Graph), and comparison with $r$.}
\end{table*}

\subsection*{Reciprocity rankings classify weighted networks}

\indent We now carry out an empirical analysis of several real weighted networks using our definitions introduced above.
We start with the global quantities $r$ and $\rho_{NM}$ defined in eqs.(\ref{eq:r}) and (\ref{eq:rho}).
In Table 1 we report the analysis of 70 biological, social and economic networks \cite{pajek2,Social1,Social2,Social3,Social4,Social5,Social6,wtw,interbank,neural,pajek}. 

All networks display a nontrivial weighted reciprocity structure  (i.e. $\rho\ne 0$), which differs from that predicted by the 3 null models considered (WCM, BCM and WRG: see Methods and Appendix). This means that the imposed constraints cannot account for the observed reciprocity.
Remarkably, we also find that networks of the same type systematically display similar values of $\rho$: for a given choice of the null model, the resulting reciprocity ranking provides a consistent (non-overlapping) classification of networks. 
However, different null models provide different estimates of reciprocity and rank the same networks differently. Some networks (social networks \cite{Social1,Social2,Social3,Social4,Social5,Social6} and the World Trade Web \cite{wtw}) always show a positive reciprocity, while others (foodwebs \cite{pajek}) always show a negative reciprocity, irrespective of the null model. 
However, other networks (interbank networks \cite{interbank}) are classified as weakly but positively reciprocal under the WCM, but as strongly negatively reciprocal under the BCM and the WRG. 
In one case (neural network \cite{neural}), the estimated level of reciprocity can be slightly positive, negative, or even consistent with zero depending on the null model.
As a consequence, the 5 interbank networks are more reciprocal than the neural network under the WCM, while the ranking is inverted under the BCM and the WRG.
Since the WCM is the most conservative model, preserving most information from empirical data, we choose to rank the networks in the Table using $\rho_{WCM}$. 

Importantly, we find that all weighted rankings are quite different from the binary analysis-based ranking \cite{myreciprocity}. While the various snapshots of the World Trade Web are systematically found to be strongly and sometimes almost perfectly reciprocal in the binary case ($0.68\le\rho_{RG}\le 0.95$ under the binary Random Graph model \cite{myreciprocity}), here we find them to be less reciprocal than social networks if the additional weighted information is taken into account.
Also, while the neural network of \emph{C. elegans} has a strong binary reciprocity ($\rho_{RG}=0.41$ \cite{myreciprocity}), here we find it to have a very weak (under the WCM), consistent with zero (under the WRG), or even negative (under the BCM) weighted reciprocity.
These important differences show that the reciprocity of weighted networks is nontrivial and irreducible to a binary description.

\begin{figure}[t!]
\includegraphics[width=.47\textwidth]{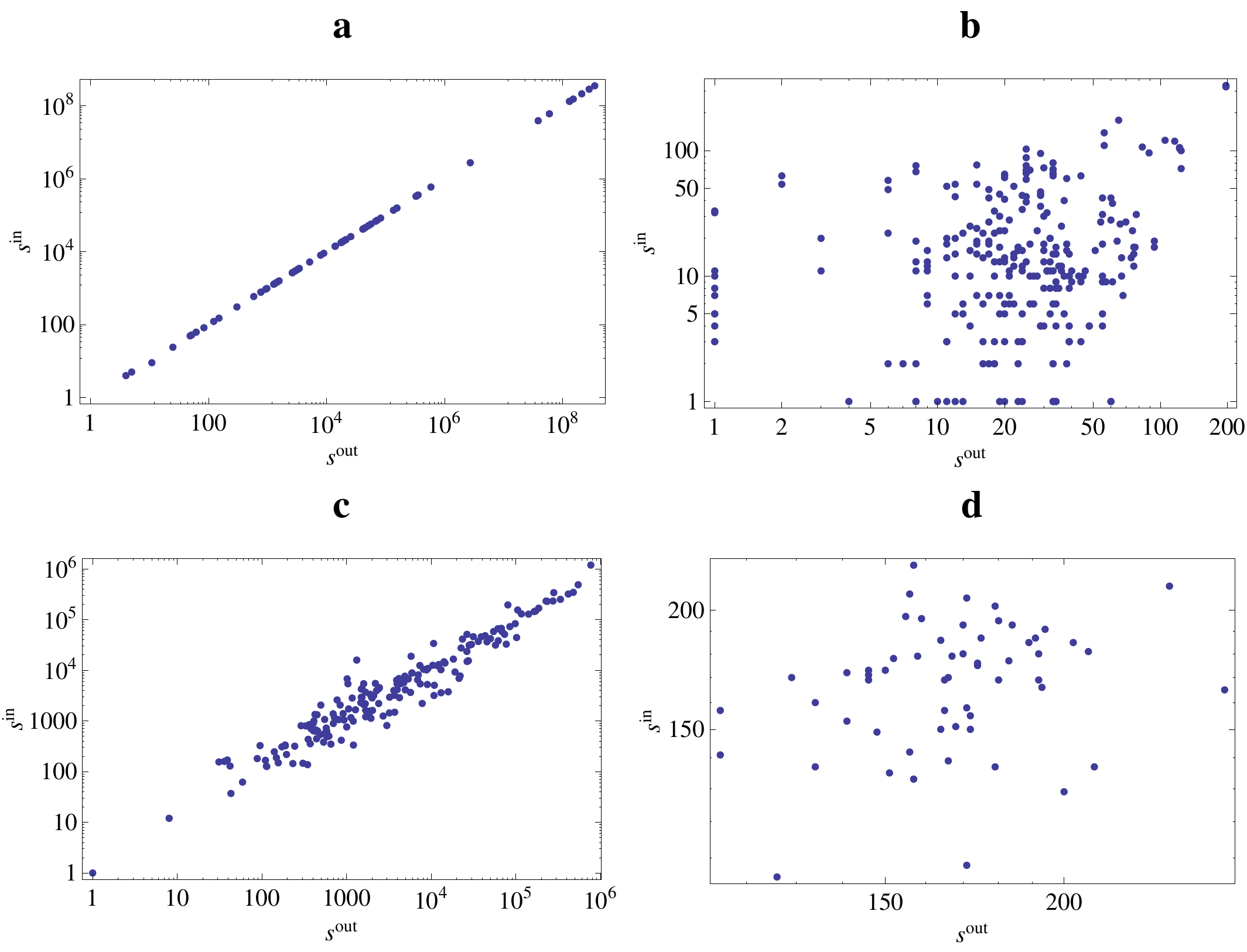}
\caption{In-strength $s^{in}_i$ versus out-strength $s^{out}_i$ in four weighted networks in increasing order of reciprocity $\rho$: a) the Everglades Marshes foodweb, b) the neural network of \emph{C. elegans}, c) the World Trade Web in the year 2000, and d)  the social network of a fraternity at West Virginia College (note that the increase in reciprocity is not necessarily associated with an increase in symmetry).}
\end{figure}

\subsection*{The role of node imbalances}

The two differences between the WCM and the WRG (see Methods and Appendix) are node imbalance ($\langle s_i^{in}\rangle$ and $\langle s_i^{out}\rangle$ are equal in the WRG and different in the WCM) and node heterogeneity (the expected strenghts of all vertices are equal in the WRG, and broadly distributed in the WCM). 
We can use the BCM as an intermediate model in order to disentangle the role of these two differences in producing the observed deviations between $\rho_{WCM}$ and $\rho_{WRG}$. 
The BCM preserves node heterogeneity but assumes node balance by regarding the observed difference between the in- and out-strength of each vertex as a statistical fluctuation around a balanced average (see Appendix).
As we show in Fig. 2, some real networks (such as foodwebs and the World Trade Web) indeed appear to display very small fluctuations around this type of node balance.
In foodwebs, where edges represent stationary flows of energy among species, the almost perfect balance is due to an approximate biomass or energy conservation at each vertex. 
In the World Trade Web, where edges represent the amount of trade among world countries, the approximate balance of vertex flows is due to the fact that countries tend to minimize the difference between their total import and their total exports, i.e. they try to `balance their payments' \cite{feenstra}.

As we show in the Appendix, the balance of vertex flows implies that,  even without introducing a tendency to reciprocate, the expected mutual weights are equal: $\langle w_{ij}\rangle=\langle w_{ji}\rangle$.
This implies a larger expected reciprocated weight $\langle W^{\leftrightarrow}\rangle$ in the BCM than in the WCM, so that $\rho_{WCM}>\rho_{BCM}$, as confirmed by Table 1. 
However, we find that $\rho_{BCM}$ and $\rho_{WRG}$ are always very similar, while they can be very different from $\rho_{WCM}$. 
This means that node imbalances, even when very weak, can have a major effect on the expected level of reciprocity. Surprisingly, we find that this effect is much stronger than that of the strikingly more pronounced node heterogeneity. 
Correctly filtering out the effects of flux balances or other symmetries can lead to counter-intuitive results: the most reciprocal of the four networks (the social network, see Table 1) is one of the least symmetric ones (see Fig. 2d), whereas the least reciprocal of the four networks (the foodweb, see Table 1) is the most symmetric one (see Fig. 2a).

\subsection*{Time evolution and fluctuations}

Since $\rho$ consistently ranks the reciprocity of networks with different properties, it can also track the evolution of reciprocity in a network that changes over time.
For this reason, in our dataset we have included 53 yearly snapshots of the World Trade Web, from year 1948 to 2000 \cite{wtw,myalessandria}.
In Fig. 3 we show the evolution of $r$, $\langle r\rangle$ and $\rho$ under the three null models. 
The plots confirm that, unlike $\rho$, $r$ is not an adequate indicator of the evolution of reciprocity, since the baseline expected value $\langle r\rangle$ (under every null model) also changes in time as a sort of moving target (Fig. 3a).

\begin{figure}
\includegraphics[width=.5\textwidth]{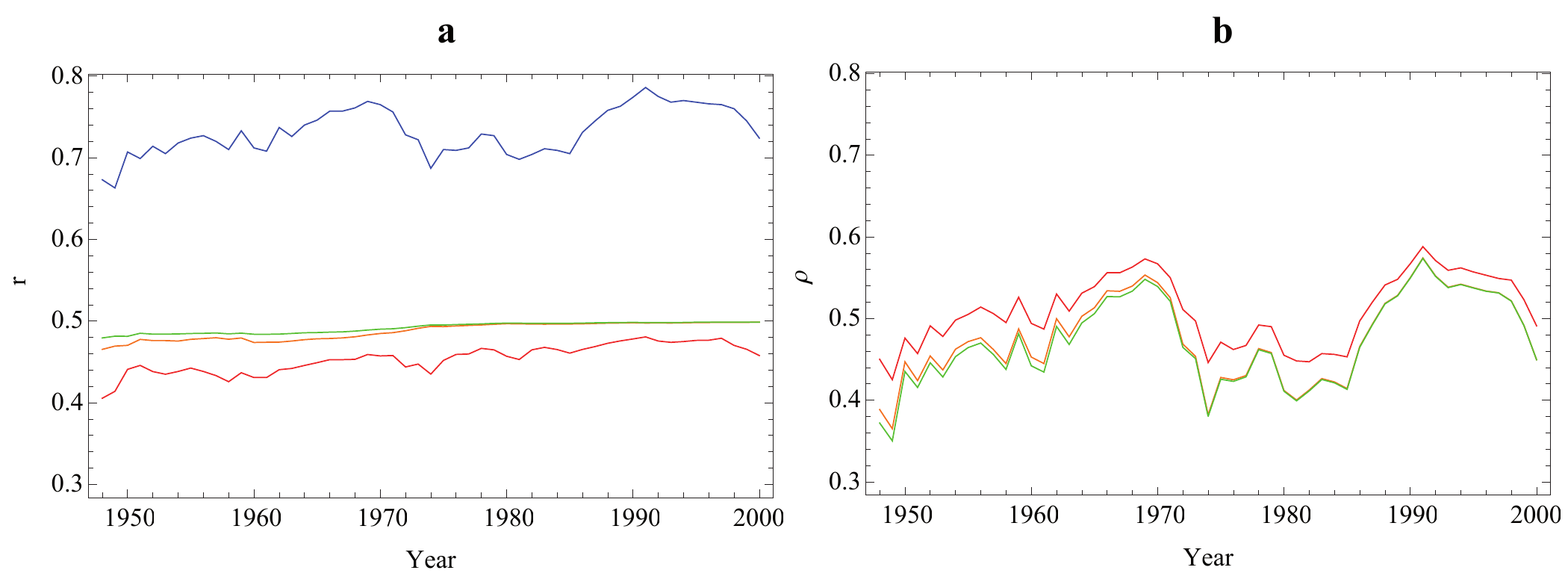}
\caption{Temporal evolution of the reciprocity of the World Trade Web during the 53 years from 1948 to 2000: a) observed value of $r$ (blue) and its expected values $\langle r\rangle_{NM}$ under the Weighted Configuration Model (red), the Balanced Configuration Model (green), and the Weighted Random Graph (orange); b) evolution of $\rho_{NM}$ under the same 3 null models as above.}
\end{figure}

Note that $\langle r\rangle_{WCM}$ fluctuates much more than $\langle r\rangle_{WRG}$ and $\langle r\rangle_{BCM}$, and its fluctuations resemble those of the observed value $r$ (see Fig. 3a). 
This is due to the fact that, while all snapshots of the network are characterized by `static' fluctuations of the empirical strengths of vertices around the balanced flux condition $s_i^{in}= s_i^{out}$ (like those in shown in Fig. 2c for the year 2000), these fluctuations have different entities in different years. Changes in the size of `static' fluctuations produce the `temporal' fluctuations observed in the evolution of $\langle r\rangle_{WCM}$, and partly also in the observed value $r$, confirming the important role of node (im)balances. 
After controlling for the time-varying entity of node imbalances (using the WCM), we indeed find that the fluctuations of $\rho_{WCM}$ are less pronounced than those of $\rho_{BCM}$ and $\rho_{WRG}$ (see Fig. 3b). 
However, the fluctuations of $r$ and $\langle r\rangle_{WCM}$ do not cancel out completely, and their resulting net effect (the trend of $\rho_{WCM}$) is still significant, indicating the strongest level of reciprocity across the three null models.

While a binary analysis of the WTW \cite{myalessandria,mysymmetry2} detected an almost monotonic increase of the reciprocity, with a marked acceleration in the 90's, we find that the weighted reciprocity has instead undergone a rapid decrease over the same decade: this counter-intuitive result confirms that the information conveyed by a weighted analysis of reciprocity is nontrivial and irreducible to the binary picture.

\subsection*{Local reciprocity structure}

\indent We now focus on the reciprocity structure at the local level of vertices, i.e. on the reciprocated and non-reciprocated strength $s^{\leftrightarrow}_i$, $s^{\leftarrow}_i$ and $s^{\rightarrow}_i$ defined in eqs.(\ref{eq:slr}-\ref{eq:sl}).
As clear from eq.(\ref{eq:wrec}), this allows us to analyse how different vertices contribute to the overall value of $W^{\leftrightarrow}$ and hence to $r$.
In order to assess whether the vertex-specific reciprocity structure is significant, rather than merely a consequence of the local topological properties of vertices, we compare the observed value of $s^{\leftrightarrow}_i$, $s^{\leftarrow}_i$ and $s^{\rightarrow}_i$ with their expected values under the WCM and the BCM. Unlike the WRG, these models preserve the total strength $s^{tot}_i= s^{out}_i+s^{in}_i$ of each vertex, thus filtering out the effects of the observed heterogenity of vertices.
In Fig. 4 we show the observed and expected values of the (non-)reciprocated strength versus the total strength $s^{tot}_i$ for the four networks already shown in Fig. 2 in order of increasing reciprocity. 

\begin{figure}
\includegraphics[width=.47\textwidth]{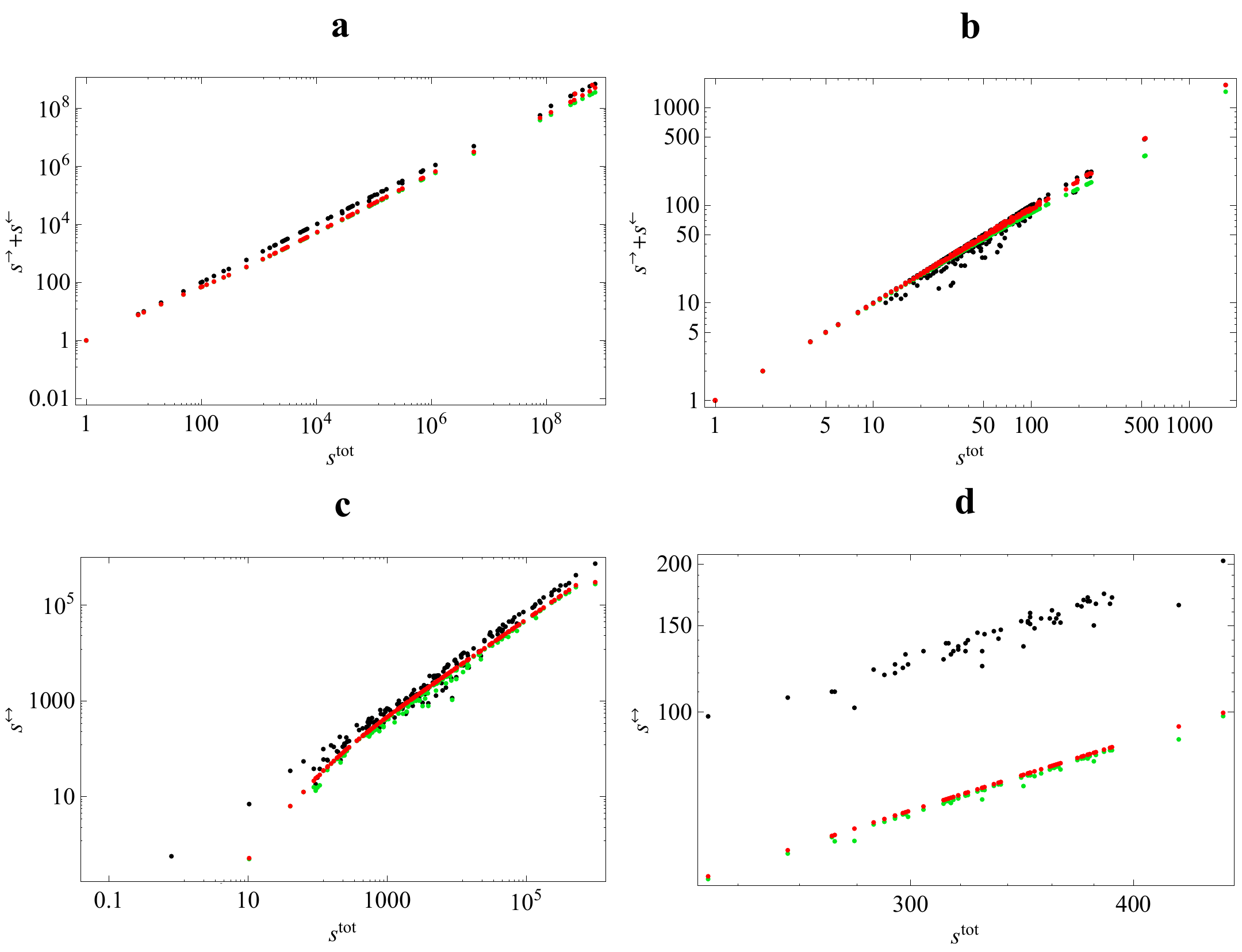}
\caption{Relationship between total ($s^{tot}_i$) and reciprocated ($s^{\leftrightarrow}_i$) or non-reciprocated ($s^{\leftarrow}_i+s^{\rightarrow}_i$) strength in four weighted networks in increasing order of reciprocity $\rho$: a) the Everglades Marshes foodweb, b) the neural network of \emph{C. elegans}, c) the World Trade Web in the year 2000, and d)  the social network of a fraternity at West Virginia College (black: real data, green: Weighted Configuration Model, red: Balanced Configuration Model).}
\end{figure}

For the anti-reciprocal networks with $\rho< 0$ (the foodweb and, under some null model, the neural network), the dominant and less fluctuating contribution to $s^{tot}_i$ comes from the non-reciprocated strength, and therefore we choose to plot $s^{\leftarrow}_i+s^{\rightarrow}_i$ versus $s^{tot}_i$ (Fig. 4a-b). 
Conversely, for the positively reciprocal networks with $\rho>0$ (the World Trade Web and the social network) the dominant contribution comes from the reciprocated strength, so we consider  $s^{\leftrightarrow}_i$ versus $s^{tot}_i$ (Fig.4c-d). 

We found very rich and diverse patterns. 
In all networks, the selected quantity displays an approximately monotonic increase with $s^{tot}_i$. 
Qualitatively, this increasing trend is also reproduced by the two null models.
However, we systematically find large differences between the latter and real data.
In the foodweb (Fig. 4a), the observed values of the \emph{non-reciprocated strength}  $s^{\leftarrow}_i+s^{\rightarrow}_i$ are always larger than the expected values (note that the separation between the two trends is exponentially larger than it appears in a log-log plot).
This shows that each vertex contributes, roughly proportionally to its total strength, to the overall anti-reciprocity of this network ($W^{\leftrightarrow}<\langle W^{\leftrightarrow}\rangle_{NM}$ and hence $\rho_{NM}<0$, see Table 1).
By contrast, in the neural network (Fig. 4b) some vertices (mostly, but not uniquely those with large $s^{tot}_i$) have a larger non-reciprocated strength than expected under the null models, while for other vertices (mostly those with small $s^{tot}_i$) the opposite is true.
This shows that the weak (and nearly consistent with zero, see Table 1) overall reciprocity of this network is the result of several opposite contributions of different vertices, that cancel each other almost completely.
The World Trade Web (Fig. 4c) also shows a combination of deviations in both directions, even if in this case for the vast majority of vertices the observed \emph{reciprocated} strength is larger than the expected one. This results in the overall positive reciprocity of the network, but again in a such a way that the global information is not reflected equally into the local one.
Finally, the social network (Fig. 4d) displays a behaviour analogous, but opposite, to that of the foodweb: the observed \emph{reciprocated} strength of each vertex systematically exceeds its expected value and gives a proportional contribution to the overall positive reciprocity.

Note that, while the striking similarity between the predictions of the WCM and the BCM in the foodweb and in the World Trade Web is not surprising, because of the very close node-balance relationship $s_{i}^{out}\approx s_{i}^{in}$ in these two networks (see Fig. 2a and 2c), in the neural network and in the social network the similarity between the predictions of the two null models is nontrivial, since node balance is strongly violated in these cases (see Fig. 2b and 2d).
\newline
\newline
\indent Having shown that the reciprocity of real weighted networks is very pronounced, we conclude our study by introducing a class of models aimed at correctly reproducing the observed patterns.
To this end, rather than proposing untestable models of network formation, we expand the null models we have considered above by enforcing additional or alternative constraints on the reciprocity structure.
This approach leads us to define the weighted counterparts of the binary Exponential Random Graphs (or $p^*$ models) with reciprocity \cite{HL,WF} and their generalizations \cite{myreciprocity,mygrandcanonical,mysymmetry2}. 
We first define three models that exactly reproduce, besides the observed heterogeneity of the strength of vertices, the observed global level of reciprocity (i.e. such that $W^{\leftrightarrow}=\langle W^{\leftrightarrow}\rangle$ and $W=\langle W\rangle$, implying $\rho=0$). Our aim is to check whether this is enough in order to reproduce the more detailed, local reciprocity structure. 

In the first model (`Weighted Reciprocity Model', see Appendix), the constraints are $s^{in}_i$ and $s^{out}_i$ for each vertex (as in the WCM), and additionally $W^{\leftrightarrow}$. This model is the analogue of the binary reciprocity model by Holland and Leinhardt \cite{HL} and replicates the overall reciprocity $r$ exactly. However, as we discuss in the Appendix, it is best suited to reproduce networks that are anti-reciprocal or, more precisely, less reciprocal than the WCM ($\rho_{WCM}<0$). Therefore, in our analysis we can only apply it to the foodwebs.
In Fig. 5a we show our results on the Everglades web.
For the sake of comparison with Fig. 4a, we plot $s^{\leftarrow}_i+s^{\rightarrow}_i$ as a function of $s^{tot}_i$.
We find that, quite surprisingly, the model does not significantly improve the accordance between real and expected trends produced by the WCM and BCM (see Fig.4a). The only difference with respect to the latter is that now a few vertices with very large $s^{tot}_i$ lie below the expected trend, while all the other vertices continue to lie above it (Fig. 5a) producing an overall $\rho=0$: so, even if all vertices appeared to contribute evenly and proportionally to the global anti-reciprocity (see Fig. 4a), adding the latter as an overall constraint is not enough in order to capture the local reciprocity structure.

\begin{figure}
\includegraphics[width=.47\textwidth]{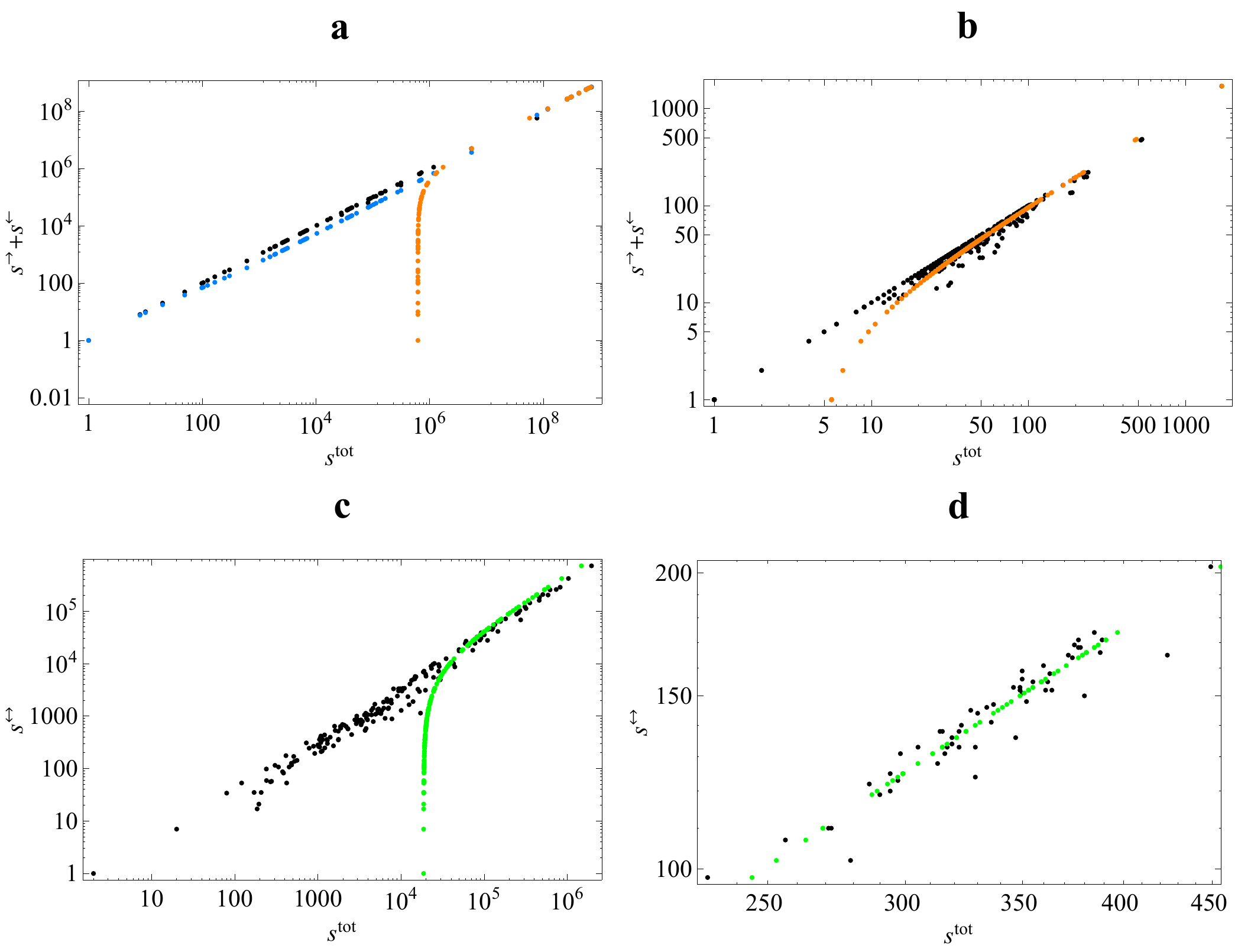}
\caption{Relationship between total ($s^{tot}_i$) and reciprocated ($s^{\leftrightarrow}_i$) or non-reciprocated ($s^{\leftarrow}_i+s^{\rightarrow}_i$) strength in four weighted networks in increasing order of reciprocity $\rho$: a) the Everglades Marshes foodweb, b) the neural network of \emph{C. elegans}, c) the World Trade Web in the year 2000, and d)  the social network of a fraternity at West Virginia College (black: real data, blue: the Weighted Reciprocity Model, orange: the Non-reciprocated Strength Model, green: the Reciprocated Strength Model; all such models reproduce the global level of reciprocity but not necessarily the local reciprocity structure).}
\end{figure}

\indent In our second model (`Non-reciprocated Strength Model', see Appendix), the constraints are $s^{\leftarrow}_i$, $s^{\rightarrow}_i$ (for each vertex), and $W^{\leftrightarrow}$.
This slightly relaxed model (potentially) generates all levels of reciprocity $r$.
However, it does not automatically reproduce the in- and out-strength sequences, therefore it is only appropriate for networks where $s^{\leftarrow}_i$ and $s^{\rightarrow}_i$ are the dominant contributions to $s^{in}_i$ and $s^{out}_i$ respectively, so that specifying the former largely specifies the latter as well. So, even if now there are no mathematical restrictions, this model is again only appropriate for networks with negative reciprocity ($\rho_{WCM}<0$). 
In Fig. 5a we show the predictions of this model on the foodweb: note that, as compared to the previous model, now the quantity $s^{\leftarrow}_i+s^{\rightarrow}_i$ is exactly reproduced by construction, while $s^{tot}_i$ is not reproduced, with most vertices lying above the expected trend and a few dominating ones lying below it. So the result is even worse than before.
In Fig. 5b we also show the performance of this model on the neural network (which actually displays $\rho_{WCM}\approx 0$, even if it still has negative reciprocity under other null models, see Table 1): even if the agreement is now much better, most data continue to lie either above or below the expected curve, confirming that the reciprocated strengths cannot be simply reconciled with the total strengths.
Note however that for networks with smaller $\rho$ this model becomes more accurate, and in the limit $W^{\leftrightarrow}\to 0$ it exactly reproduces all the strength sequences of any network.

\indent Our third model (`Reciprocated Strength Model', see Appendix) is a `dual' one appropriate in the opposite regime of strong positive reciprocity (i.e. $\rho_{WCM}>0$, especially in the limit $\rho_{WCM}\to 1$). The constraints are now $s^{\leftrightarrow}_i$ (for each vertex) and the total weight $W$ (note that, as a consequence, also the non-reciprocated total weight $W^{\rightarrow}\equiv W-W^{\leftrightarrow}$ is kept fixed).
This model is most appropriate for networks where $s^{\leftrightarrow}_i$ is the dominant contribution to $s^{tot}_i$.
In Fig. 5c we show the predictions of this model on the World Trade Web. Now $s^{\leftrightarrow}_i$ is obviously always reproduced, while $s^{tot}_i$ instead is not reproduced for all vertices.
In Fig. 5d we show the results for the social network, and in this case we find that the model reproduces real data remarkably well.
This confirms that the model is particularly appropriate for strongly reciprocal networks.
We therefore find that, as in the dual case discussed above, if the overall reciprocity is moderate then the constraints are in general not enough in order to characterize the local reciprocity structure.
However, in networks with strong overall reciprocity, this model accurately (and exactly in the limit $W^{\to}\to 0$) reproduces all the local reciprocity structure.

\section*{Discussion}

The above three models produce the correct level of global reciprocity (i.e. $\langle r\rangle=r$ or $\rho=0$) but not necessarily the correct local reciprocity structure.
In networks with strong (either positive or negative) reciprocity, the local reciprocity structure can be simply inferred from the global one, plus some information about the heterogeneity of vertices (some strength sequence). Conversely, in networks with moderate reciprocity the local patterns are irreducible to any overall information, and thus constitute intrinsic heterogeneous features.
In this case, it is unavoidable to use a model that fully reproduces the three quantities $s^{\leftarrow}_i$, $s^{\rightarrow}_i$ and $s^{\leftrightarrow}_i$ separately for each vertex, by treating them as constraints.
In the Appendix we describe this model, that we denote as the Weighted Reciprocated Configuration Model (WRCM) in detail. 
Using this model, all the plots in Fig. 5 are automatically reproduced exactly, by construction.
Therefore we believe that this model represents an important starting point for future analyses of higher-order topological properties in weighted networks. In particular, we foresee two main applications.

The first application is to the analysis of weighted `motifs', i.e. the abundances of all topologically distinct subgraphs of three or four vertices \cite{motifs,foodwebmotifs}.
In the binary case, it has been realized that such subgraphs are important building blocks of large networks, and that their abundance is not trivially explained in terms of the dyadic structure.
This result can only be obtained by comparing the observed abundances with their expectation values under a null model that separately preserves the number of reciprocated and non-reciprocated (in-coming and out-going) links of each vertex.
In the weighted case, no similar analysis has been carried out so far, because of the lack of an analogous method, like the WRCM defined here, to control for the reciprocated and non-reciprocated connectivity properties separately.

The second application is to the problem of community detection \cite{santo} in weighted directed networks, i.e. the identification of densely connected modules of vertices.
Most approaches attempt to find the partition of the network that maximizes the so-called `modularity', i.e. the total difference between the observed weights of intra-community links and their expected values under the WCM. 
In networks where the observed reciprocity is not reproduced by the WCM (as all networks in the present study), the difference between observed and expected weights is not necessarily due to the presence of community structure, as it also receives a (potentially strong) contribution by the reciprocity.
This means that, in order to filter out the effects of reciprocity from community structure, in the modularity function one should replace the expected values under the WCM with the expected values under the WRCM.

The ever-increasing gap between the growth of data about weighted networks and our poor understanding of their dyadic properties led us to propose a rigorous approach to the reciprocity of weighted networks.
We showed that real networks systematically display a rich and diverse reciprocity structure, with several interesting patterns at the global and local level.
We believe that our results form an important starting point to answer many open questions about the effect of reciprocity on higher-order structural properties and on dynamical processes taking place on real weighted networks.

\section*{Methods}
Equation (\ref{eq:rho}) in the Results section introduces the quantity $\rho_{NM}$, as the normalized difference between the observed value of the weighted reciprocity $r$ and its expected value under a chosen null model $\langle r\rangle_{NM}$.
The introduction of $\rho_{NM}$ has two important consequences. Firstly, networks with different parameters can be ranked from the most to the least reciprocal using the measured value of $\rho_{NM}$. Secondly, and consequently, the reciprocity of a network that evolves in time can be tracked dynamically using $\rho_{NM}$ even if other topological properties of the network change (as is typically the case).
Clearly, the above considerations apply not only to the global quantity $r$, but also to the edge- and vertex-specific definitions we have introduced in eqs.(\ref{eq:wlr}-\ref{eq:wl}) and (\ref{eq:slr}-\ref{eq:sl}).
For this reason, in the Appendix we introduce and study three important null models in great detail. We briefly describe these models below.

\subsection*{Null models: the Weighted Random Graph model}

\indent To start with, we consider a network model with the same total weight $W$ as the real network but with no tendency towards or against reciprocation, i.e. a directed version of the Weighted Random Graph (WRG) model \cite{myWRG}. 
This allows us to quantify for the first time the baseline level of reciprocity $\langle r\rangle_{WRG}$ expected by chance in a directed network with given total weight.
However, this null model is severely limited by the fact that it is completely homogeneous in two respects (see the Appendix): it generates networks where each vertex $i$ has the same expected in- and out-strength ($\langle s_i^{in}\rangle_{WRG}=\langle s_i^{out}\rangle_{WRG}\equiv \langle s_i\rangle_{WRG}\quad \forall i$), and moreover this value is common to all vertices ($\langle s_i\rangle_{WRG}=\langle s\rangle_{WRG} \quad\forall i$).

\subsection*{Null models: the Weighted Configuration model}

\indent A popular and more appropriate null model that preserves the observed intrinsic heterogeneity of vertices is one where all vertices have the same in-strength and out-strength as in the real network, i.e. the directed Weighted Configuration Model (WCM) \cite{weightedconfiguration}. 
In such model, since $\langle s_i^{in}\rangle_{WCM}=s_i^{in}$ and $\langle s_i^{out}\rangle_{WCM}=s_i^{out}\quad\forall i$, the two sources of homogeneity characterizing the WRG are both absent: each vertex has different values of the in-strength and out-strength, and these values are also heterogeneously distributed across vertices.
In other words, this model preserves the in- and out-strength sequences separately.

\subsection*{Null models: the Balanced Configuration model}

\indent Another important null model that we introduce here for the first time is one that allows us to conclude whether the observed asymmetry of fluxes is consistent with a fluctuation around a balanced network (i.e. one where the net flow at each vertex is zero). This model, that we denote as the Balanced Configuration Model (BCM), is somewhat intermediate between the above two models, as it assumes (like the WRG) that the expected in- and out-strength of each vertex are the same, i.e. that the two observed values $s_i^{in}$ and $s_i^{out}$ are fluctuations around a common expected value $\langle s_i\rangle_{BCM}=(s_i^{in}+s_i^{out})/2$, but at the same time preserves (as the WCM) the strong heterogeneity of vertices (i.e. in general $\langle s_i\rangle_{BCM}\ne \langle s_j\rangle_{BCM}$ if $i\ne j$). This model preserves the total strength $s_i^{tot}\equiv s_i^{in}+s_i^{out}$ of each vertex, but not the in- and out-strength separately. \\

Note that all the above null models preserve the total weight of the original network, i.e. $\langle W\rangle_{NM}=W$. However, they do not automatically preserve the reciprocity (neither locally nor globally). Our aim is to understand whether the observed reciprocity can be simply reproduced by one of the null models (and is therefore trivial), or whether it deviates systematically from the null expectations. 
In the next section we show that the latter is true, and that the reciprocity structure is a robust and novel pattern characterizing weighted networks.

\subsection*{A unifying formalism}

\indent As we show in the Appendix, it is possible to characterize all the above null models analytically, and thus to calculate the required expected values exactly. Even if the final expressions are rather simple, their derivation is in some cases quite involved and requires further developments of mathematical results that have appeared relatively recently in the literature \cite{newman_expo,mybosefermi,myWRG}. 
Moreover, the crucial step that fixes the values of the parameters of all models requires the application of a maximum-likelihood method that has been proposed by two of us only recently \cite{mymethod}. 
It is for the above reasons, we believe, that the reciprocity of weighted networks has not been studied as intensively as its binary counterpart so far. 
By putting all the pieces together, we are finally able to approach the problem in a consistent and rigorous way.
Importantly, the framework wherein our null models are introduced (maximum-entropy ensembles of weighted networks with given properties) extends to the weighted case, and at the same time formally unifies, recent randomization approaches proposed by physicists and well-established models of social networks introduced by statisticians, i.e. the so-called Exponential Random Graphs or $p^*$ models (see the Appendix). While a variety of specifications for the latter exist in the binary graph case \cite{WF,snijders,pattison}, very few results for weighted graphs are available \cite{ciccio}. Our contribution opens the way for the introduction of more general families of Exponential Random Graphs for weighted networks.
Indeed, besides the null models discussed above, we will also introduce the first models that correctly reproduce the observed reciprocity structure, either at the global (but not necessarily local) level, or at the local (and consequently also global) level.
It is worth mentioning that our approach makes use of exact analytical expressions, and allows to find the correct values of the parameters both in the null models and in the models with reciprocity. By contrast, the common methods available in social network analysis to estimate binary Exponential Random Graphs rely on approximate techniques such as Markov Chain Monte Carlo or pseudo-likelihood approaches \cite{WF,snijders,pattison}.
Another advantage is that the method we employ allows us to obtain the expected value of any topological property mathematically, and in a time as short as that required in order to measure the same property on the original network \cite{mymethod}. Unlike other randomization approaches \cite{MS,MSZ}, we do not need to computationally generate several randomized variants of the original network and take (approximate, and generally biased \cite{bias}) sample averages over them.

Comparing real data with the above null models, and the null models among themselves, allows us to separate different sources of heterogeneity observed in networks. This is a key step towards understanding the origin of the reciprocity structure of real weighted networks. 

\begin{acknowledgments}
 D. G. acknowledges support from the Dutch Econophysics Foundation (Stichting Econophysics, Leiden, the Netherlands) with funds from beneficiaries of Duyfken Trading Knowledge BV, Amsterdam, the Netherlands.

F. R. acknowledges support from the FESSUD project on ``Financialisation, economy, society and sustainable development'', Seventh Framework Programme, EU.
\end{acknowledgments}





\section*{appendix}

\section{Reciprocity of binary networks}
Before considering the reciprocity of weighted networks, we briefly recall the basic definitions in the binary case, that were originally introduced to describe the mutual relations taking place between vertex pairs \cite{WF,HL}.

\subsection{Reciprocity as the fraction of bidirectional links}

For binary, directed networks the reciprocity is defined as the fraction of links having a ``partner'' pointing in the opposite direction:

\begin{equation}
r^b\equiv\frac{L^\leftrightarrow}{L}
\label{r}
\end{equation}

\noindent where $L=\sum_{i\ne j}a_{ij}$ and $L^\leftrightarrow=\sum_{i\ne j}a_{ij}a_{ji}$. The above quantity, $r^b$, is not independent on the link density (or connectance) $c\equiv\frac{L}{N(N-1)}=\frac{\sum_{i\ne j}a_{ij}}{N(N-1)}\equiv\bar{a}$: on the contrary, it can be shown that $c$ is the expected value of $r^b$ under the Directed Random Graph Model (DRG in what follows) \cite{mygrandcanonical,myreciprocity}. 
In the DRG, a directed link is placed with probability $p$ between any two vertices, i.e. $\langle a_{ij}\rangle_{DRG}=p,\:\forall\:i, j$ (with $i\neq j$). This implies

\begin{equation}
\langle r^b\rangle_{DRG}\equiv\frac{\langle L^{\leftrightarrow}\rangle}{\langle L\rangle}=\frac{N(N-1)p^{2}}{N(N-1)p}=p\equiv\frac{L}{N(N-1)}=c
\label{r2}
\end{equation}

\noindent showing that the expected value of $r^b$ coincides with the fundamental parameter of this null model, and hence depends on $L$ and $N$.
In order to assess whether there is positive or negative reciprocity, one should compare the measured $r^b$ with its expected value $\langle r^b\rangle_{DRG}$.
This means that $r^b$ cannot be used to consistently rank networks with different values of $L$ and $N$, because they have different reference values. Also, and consequently, $r^b$ cannot be used to track the evolution of a network that changes in time, because $L$ and/or $N$ will also change \cite{myreciprocity}.

\subsection{Reciprocity as a correlation coefficient}

This is why a different definition of reciprocity was proposed \cite{myreciprocity}, trying to control for the time-varying properties by means of the Pearson correlation coefficient between the transpose elements of the adjacency matrix  \cite{mysymmetry2}:

\begin{equation} 
\rho^b\equiv\frac{\sum_{i\neq j}(a_{ij}-c)(a_{ji}-c)}{\sum_{i\ne j}(a_{ij}-c)^2}=\frac{r^b-c}{1-c}=\frac{r^b-\langle r^b\rangle_{DRG}}{1-\langle r^b\rangle_{DRG}}.
\label{rho}
\end{equation}

A symmetrical adjacency matrix (as those for binary, undirected networks) represents a network with the highest values of $r^b$ and $\rho$ (both equal to 1), whereas a fully asymmetrical one, with $zero$ values mirroring $unit$ values on opposite sides of the main diagonal (like a triangular matrix), displays the lowest value, being $r^b=0$ and $\rho=-c/(1-c)$) \cite{myreciprocity}.
This meaningful definition of reciprocity automatically discounts density effects, i.e. the expectation value of $r^b$ (under the DRG). As a result, consistent rankings and temporal analyses become possible in terms of $\rho$.

\section{Reciprocity of weighted networks}

In what follows we provide additional information about the possible generalization of the reciprocity to the weighted case.

\subsection{From binary to weighted\label{sec2}: the first route}

By looking at eq.(\ref{rho}), it is not clear whether a generalization to the weighted case should start from the first term on the left (i.e. as a correlation coefficient) or from the last term on the right (i.e. as the normalized excess from a random expectation).
This ambiguity comes from the fact that, for weighted networks, those two terms are no longer equivalent (as we now show).
We therefore start by attempting the first route, and then consider the second one.

If we follow the binary recipe from left to right, we define the weighted reciprocity as the Pearson correlation coefficient (where, as usual, $\bar{w}=\frac{\sum_{i\neq j}w_{ij}}{N(N-1)}=\frac{W_{tot}}{N(N-1)}$). After some algebra, this implies

\begin{equation} 
\rho\equiv\frac{\sum_{i\neq j}(w_{ij}-\bar{w})(w_{ji}-\bar{w})}{\sum_{i\ne j}(w_{ij}-\bar{w})^2}=\frac{r-c^w}{1-c^w}
\label{rhow}
\end{equation}

\noindent where, in order to produce a result formally equivalent  to eq.(\ref{rho}), we have defined the weighted analogues of $r$ and $c$ as follows:

\begin{equation} 
r\equiv\frac{\sum_{i\neq j}w_{ij}w_{ji}}{\sum_{i\neq j}w_{ij}^2},\:c^w\equiv\frac{\bar{w}^2}{\sum_{i\neq j}w_{ij}^2/N(N-1)}
\label{rcw}
\end{equation}

\noindent Note that the equivalence $\bar{a}=c$, valid for the binary case, no longer holds: $\bar{w}\neq c^w$. The previous expressions generalize the binary ones and reduce to them when substituting the $a_{ij}$'s in place of the $w_{ij}$'s. Moreover, interestingly enough, the coefficient $c^w$ can be expressed as a function of the weights' distribution mean, $m$, and standard deviation, $s$, or, in an equivalent way, as a function of the so-called coefficient of variation, $c_{v}=s/m$, as

\begin{equation} 
c^w=\frac{m^2}{m^2+s^2}=\frac{1}{1+c_{v}^2}.
\label{cw}
\end{equation}

We could be tempted to interpret $c^w$ as the weighted counterpart of the binary connectance and, $r$ as the weighted counterpart of eq.(\ref{r}). However, we can show a simple case for which the above ``product-over-squares'' definition above fails in measuring our intuitive notion of reciprocity. Let us consider a simple network like that in Fig. 1. 

\begin{figure}
\includegraphics[scale=0.5]{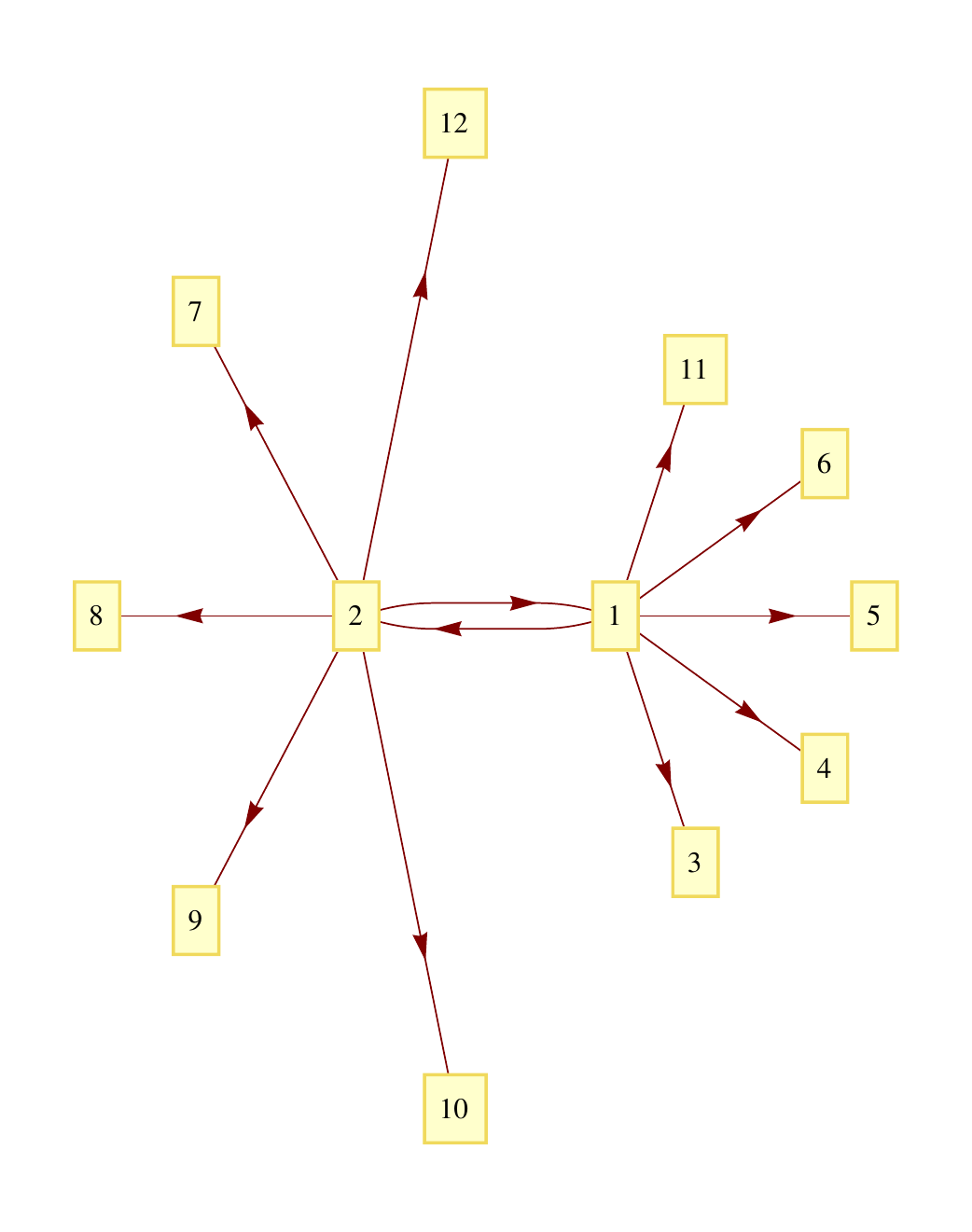}
\caption{A double-star network, with only one reciprocated pair of weights.}
\label{fig:POS}
\end{figure}


If we calculate $r$ by choosing $w_{12}=w_{21}$, we obtain

\begin{equation} 
r_{1}=\frac{2w_{12}^2}{2w_{12}^2+\sum_{i,j\neq (1,2), (2,1)}w_{ij}^2}
\label{pos_ex}
\end{equation}

\noindent where the sum in the denominator includes all the weights different from the central ones. Now, let us imagine a second situation where $w_{21}=w_{12}+1$; the calculations, now, would give

\begin{equation} 
r_{2}=\frac{2w_{12}(w_{12}+1)}{w_{12}^2+(w_{12}+1)^2+\sum_{i,j\neq (1,2), (2,1)}w_{ij}^2}
\label{pos_ex2}
\end{equation}

\noindent and we would intuitively require that $r_{2}<r_{1}$, for every choice of the involved weights, because of the greater disparity between the two central flows. However, it can be shown that under certain circumstances exactly the opposite result is obtained, by simply changing the non-central weights. In fact, by choosing the latter to satisfy the condition 

\begin{equation} 
\sum_{i,j\neq (1,2), (2,1)}w_{ij}^2>w_{12}
\label{delta}
\end{equation}

\noindent the very counter-intuitive result $r_{2}-r_{1}>0$ is obtained. This shows that eq.(\ref{rhow}) is not a good choice for a weighted extension of eq.(\ref{rho}).

Before considering the alternative route, we observe that we could also imagine to define a slightly different correlation coefficient, only between the two triangular blocks of the weighted adjacency matrix: the upper-diagonal one and the lower-diagonal one. This would be defined as

\begin{equation} 
\rho^{'}\equiv\frac{\sum_{i<j}(w_{ij}-\bar{w}_{u})(w_{ji}-\bar{w}_{l})}{\sqrt{\sum_{i<j}(w_{ij}-\bar{w}_{u})^2\sum_{i<j}(w_{ji}-\bar{w}_{l})^2}}
\end{equation}

\noindent where $\bar{w}_{u}\equiv\frac{\sum_{i>j}w_{ij}}{N(N-1)}$ is the upper-diagonal mean and $\bar{w}_{l}\equiv\frac{\sum_{i>j}w_{ji}}{N(N-1)}$ is the lower-diagonal mean. 
Again, this definition has an undesirable performance. This is evident if we imagine a matrix whose transposed entries are defined as $w_{ij}$ and $w_{ji}\equiv \lambda w_{ij}$ (with $i<j$). In this case, we would have

\begin{equation} 
\rho^{'}=\frac{\sum_{i<j}(w_{ij}-\bar{w}_{u})(\lambda w_{ij}-\lambda \bar{w}_{u})}{\sqrt{\sum_{i<j}(w_{ij}-\bar{w}_{u})^2\sum_{i<j}(\lambda w_{ij}-\lambda\bar{w}_{u})^2}}=1
\end{equation}

\noindent independently of the value of $\lambda$! So we could arbitrarily rise or lower the value of $\lambda$, thus making the matrix more and more asymmetric, without measuring this effect at all.
Note that this circumstance is impossible in the binary case, as all weights are forced to be either zero or one, and therefore the only allowed value for $\lambda$ is one. 

The two examples above show that correlation-based definitions of reciprocity, while having a satisfactory behaviour in the binary case, become problematic in the weighted one.
Unfortunately, the few attempts that have been proposed so far in order to characterize the reciprocity of weighted networks \cite{jari,fagiolo,achen,faloutsos} are all based on measures of correlation or symmetry between mutual weights. 
Later, we show that symmetry-based measures are also flawed. 
Together with our results above, this means that all the available measures fail in providing a consistent and interpretable characterizaton of the reciprocity of weighted networks. 

\subsection{From binary to weighted\label{sec2}: the second route}

We now consider the second route, i.e. a definition that starts from generalizing the last term in eq.(\ref{rho}).
This means that we are now free to first generalize $r$ in a satisfactory way, rather than as a forced effect of the correlation-based definition, and then calculate its expected value under some appropriate null model. 
To this end, we note that the binary nature of the variables defining $r^b$ allows us to rewrite it in a very suggestive way:

\begin{equation}
r^b\equiv\frac{L^{\leftrightarrow}}{L}=\frac{\sum_{i\neq j}a_{ij}a_{ji}}{\sum_{i\neq j}a_{ij}}=\frac{\sum_{i\neq j}\min[a_{ij},\:a_{ji}]}{\sum_{i\neq j}a_{ij}}.
\end{equation}

The previous relation is consistent with the intuitive meaning of reciprocity, as a measure of the quantity of mutually-exchanged flux between vertices. So we can extend this definition to the weighted case, to obtain

\begin{equation}
r\equiv\frac{W^{\leftrightarrow}}{W}=\frac{\sum_{i\neq j}\min[w_{ij},\:w_{ji}]}{\sum_{i\neq j}w_{ij}}.
\label{rmin}
\end{equation}
where we have defined the \emph{total reciprocated weight} as $W^{\leftrightarrow}\equiv\sum_{i\neq j}\min[w_{ij},\:w_{ji}]$.
This definition does not suffer from the same limitations of the previous one. On the contrary, the more the difference between mutual links, the less the reciprocity, because the numerator would not change, while the denominator would become larger. Note that $r\le 1$: in fact, since we are considering pairs of nodes at a time, we can rewrite it as

\begin{equation}
r=\frac{\sum_{i<j}\left(\min[w_{ij},\:w_{ji}]+\min[w_{ij},\:w_{ji}]\right)}{\sum_{i<j}\left(\min[w_{ij},\:w_{ji}]+\max[w_{ij},\:w_{ji}]\right)}.
\label{rmin2}
\end{equation}

Another advantage of this second definition is the possibility of mutuating from it the concept of \emph{reciprocated strength} in the same way as the concept of reciprocated degree was defined:

\begin{equation}
k^{\leftrightarrow}_{i}\equiv\sum_{j(\neq i)}a_{ij}a_{ji}\quad\rightarrow\quad s^{\leftrightarrow}_{i}\equiv\sum_{j(\neq i)}\min[w_{ij},\:w_{ji}]
\label{srec}
\end{equation}

\noindent so that a very impressive definition of reciprocity can be given, as

\begin{equation}
r^b=\frac{\sum_{i}k^{\leftrightarrow}_{i}}{L}\quad\rightarrow\quad r=\frac{\sum_{i}s^{\leftrightarrow}_{i}}{W}.
\label{srec2}
\end{equation}

A further feature of this quantity is its scale-invariance: if all the weights are multiplied by a scale factor, $w_{ij}\rightarrow\lambda w_{ij}$, $r$ does not change, as shown below:

\begin{equation}
r_{\lambda}=\frac{\sum_{i\neq j}\min[\lambda w_{ij},\:\lambda w_{ji}]}{\sum_{i\neq j}\lambda w_{ij}}=\frac{\lambda\sum_{i\neq j}\min[w_{ij},\:w_{ji}]}{\lambda\sum_{i\neq j}w_{ij}}=r.
\end{equation}

Moreover, in the case we had a matrix with transposed entries defined as $w_{ij}$ and $w_{ji}\equiv\lambda w_{ij}$ (with $i<j$) as in the example considered before, we would find

\begin{equation}
r=\left\{ \begin{array}{ll}
\frac{\sum_{i<j}2w_{ij}}{\sum_{i<j}(\lambda+1)w_{ij}}=\frac{2}{(\lambda+1)}, & \textrm{if $\lambda>1$}\\
\frac{\sum_{i<j}2w_{ij}}{\sum_{i<j}2w_{ij}}=1, & \textrm{if $\lambda=1$}\\
\frac{\sum_{i<j}2\lambda w_{ij}}{\sum_{i<j}(\lambda+1)w_{ij}}=\frac{2\lambda}{(\lambda+1)}, & \textrm{if $\lambda<1$}
\end{array} \right.
\end{equation}

\noindent thus obtaining a continuous function with a global maximum in $\lambda=1$ as it should be (see Fig. 2). 

\begin{figure}
\includegraphics[scale=0.6]{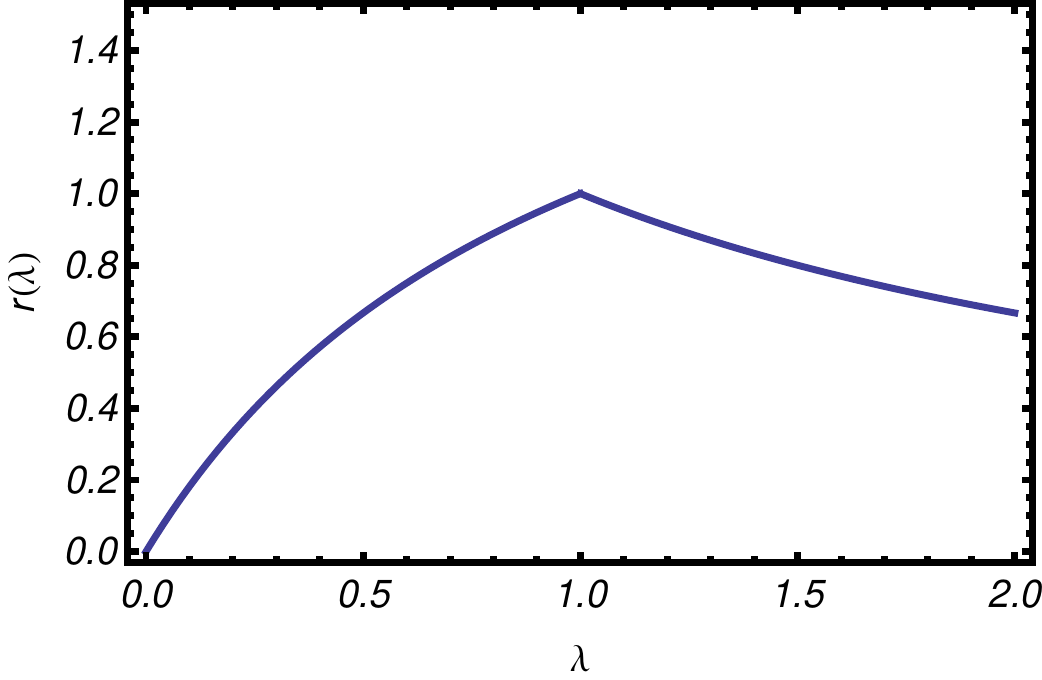}
\caption{The quantity $r$ as a function of $\lambda$.}
\label{fig:rec_l}
\end{figure}


It follows that the appropriate weighted generalization of eq.(\ref{rho}) is 

\begin{equation}
\rho_{NM}\equiv\frac{r-\langle r\rangle_{NM}}{1-\langle r\rangle_{NM}}
\end{equation}

\noindent where $r$ is defined by eq.(\ref{rmin}) and its expected value has to be computed according to a chosen null model ($NM$). Indeed, this choice also gives us the possibility to choose different null models, and compare their effects on $\rho$. From $r\le 1$, it follows that $\rho\le 1$.

\section{Null models\label{sec3}}

In this section we describe in detail the three null models we considered in order to carry out our analysis.
We adopt the formalism of Exponential Random Graphs or $p^*$ models, which allows to obtain maximally random ensembles of networks with specified constraints.
Exponential random graphs were first introduced in social network analysis \cite{WF,HL,snijders,pattison} and then recently rephrased within a maximum-entropy approach typical of statistical physics  \cite{newman_expo}. We adopt the latter notation, as it is more practical when, rather than approaching the problem using approximate techniques such as Markov Chain Monte Carlo or pseudo-likelihood \cite{snijders,pattison}, one can solve the model analytically and obtain exact results as we do below.

Exponential Random Graphs are very useful when one needs to understand, as in our case, the expected effects of a given set of topological properties, $\vec{C}$ (such as the total weight, or the strength sequence) on the structure of networks.
Recently, a method based on the maximum-likelihood principle was proposed \cite{mymethod} in order to fit exponential random graphs to a real-world graph $\textbf{G}^*$ exactly \cite{mymethod}. This method provides null models which specify the effects of one or more constraints on the structure of the \emph{particular} network $\textbf{G}^*$, and hence allows to empirically detect patterns in the latter, identified as deviations from the model's predictions \cite{mymethod}. 
In the method, maximum-entropy exponential random graphs are generated by specifying an ensemble $\mathcal{G}$ of allowed graphs, and by looking for the probability $P(\textbf{G}|\vec{\theta})$ of generating a single graph $\textbf{G}$ in the ensemble in such a way that the Shannon entropy

\begin{equation}
S(\vec{\theta})\equiv-\sum_{\textbf{G}\in \mathcal{G}}P(\textbf{G}|\vec{\theta})\ln P(\textbf{G}|\vec{\theta})
\label{eq:entropy}
\end{equation}

\noindent is maximum, under the constraints that the probability is properly normalized, $\sum_{\textbf{G}\in\mathcal{G}}P(\textbf{G}|\vec{\theta})=1,\:\forall\vec{\theta}$, and that the expected value 

\begin{equation}
\langle \vec{C}\rangle_{\vec{\theta}}\equiv \sum_{\textbf{G}\in\mathcal{G}} \vec{C}(\textbf{G})P(\textbf{G}|\vec{\theta})
\end{equation}

\noindent of the set $\vec{C}$ of enforced topological properties equals the particular  value $\vec{C}^*\equiv \vec{C}(\textbf{G}^*)$ observed on the real network $\textbf{G}^*$:

\begin{equation}
\langle \vec{C}\rangle_{\vec{\theta}^*}=\vec{C}^*.
\label{eq:con}
\end{equation}

In the above expressions, $\vec{\theta}$ is a vector of Langrange multipliers allowing to tune the value of $\langle \vec{C}\rangle_{\vec{\theta}}$, and $\vec{\theta}^*$ is the specific value of $\vec{\theta}$ that makes $\langle \vec{C}\rangle_{\vec{\theta}}$ coincide with $\vec{C}^*$, as dictated by the maximum-likelihood principle \cite{mylikelihood}. The solution to the above constrained maximization problem is

\begin{equation}
P(\textbf{G}|\vec{\theta}^*)=\frac{e^{-H(\textbf{G}|\vec{\theta}^*)}}{Z(\vec{\theta}^*)}
\end{equation}

\noindent where

\begin{equation}
H(\textbf{G}|\vec{\theta}^*)=\vec{\theta}^*\cdot\vec{C}(\textbf{G})
\end{equation}

\noindent is sometimes called the \emph{graph Hamiltonian} and

\begin{equation}
Z(\vec{\theta}^*)=\sum_{\textbf{G}\in\mathcal{G}}{e^{-H(\textbf{G}|\vec{\theta}^*)}}
\end{equation}

\noindent is the \emph{partition function}, ensuring that the probability is properly normalized. The above formal results translate into specific quantitative expectations when a particular choice of the constraints, $\vec{C}$, is made. 

Once the numerical values of the Lagrange multipliers are found, they can be used to find the ensemble average, $\langle X\rangle^*$, of any topological property $X$ of interest:

\begin{equation}
\langle X\rangle^*=\sum_{\mathbf{G}\in\mathcal{G}}X(\mathbf{G})P(\mathbf{G}|\vec{\theta}^*).
\end{equation}

The exact computation of the expected values can be very diffcult. For this reason it is often necessary to rest on the linear approximation method even if, in what follows, the only approximation will be that of treating the expected value of a ratio, as the ratio of the expected values: $\langle n/d\rangle\simeq \langle n\rangle/\langle d\rangle$.

The next subsections will be devoted to the description of the null modes used in the main text.

\subsection{The Directed Weighted Random Graph (WRG) model}

We start with the simplest case, which is the most direct generalization of the binary, undirected random graph (Erd\H{o}s-R\'enyi) model. For an ensenble of binary, undirected networks, it was shown \cite{newman_expo} that, if the only constraint $C$ is the total number $L$ of links (i.e. $H(\textbf{G},\theta)=\theta L$), then the probability $P(\textbf{G}|\theta)$ coincides with that of the \textit{Erd\H{o}s-R\'enyi Random Graph Model}. In the latter, each pair of vertices is connected with the same probability $p$, all pairs of vertices being sampled independently of each other. In the framework of exponential random graphs, the probability $p$ is simply a function of $\theta$.

The random graph model has already been generalized to the undirected, weighted case \cite{myWRG}, by considering an ensemble of networks with non-negative, integer-valued edge weights ($w_{ij}\in\mathbf{N},\:\forall\:i, j$) and imposing, as the only constraint, the total weight, $W=\sum_{i<j}w_{ij}$. The result is the \emph{Undirected Weighted Random Graph} model \cite{myWRG}, where each pair of vertices is still independent as in its binary counterpart, and connected by an edge of weight $w$ with probability $q(w)=p^w(1-p)$, where $p\equiv e^{-\theta}$.

Here we introduce the directed version of the weighted random graph. The hamiltonian of the WRG is

\begin{equation}
H(\textbf{G}|\theta)=\theta W=\theta\sum_{i\neq j}w_{ij};
\end{equation}

\noindent thus, the partition function becomes

\begin{eqnarray}
Z(\theta)&=&\sum_{\textbf{G}\in\mathcal{G}}e^{-H(\textbf{G}|\theta)}=\sum_{\textbf{G}\in\mathcal{G}}e^{-\theta\sum_{i\neq j}w_{ij}}=\nonumber\\
&=&\prod_{i\neq j}\sum_{w_{ij}=0}^{+\infty}e^{-\theta w_{ij}}=\prod_{i\neq j}(1-e^{-\theta})^{-1}
\end{eqnarray}

\noindent (provided that $e^{-\theta}<1$), that is a product over the $N(N-1)$ independent random variables, identified with the orderd pairs of the network's $N$ nodes. So, every (non-negative, integer-valued) weighted network in the grandcanonical ensemble has the following probability

\begin{equation}
P(\textbf{G})=\frac{\prod_{i\neq j}e^{-\theta w_{ij}}}{\prod_{i\neq j}(1-e^{-\theta})^{-1}}\equiv\prod_{i\neq j}p^{w_{ij}}(1-p)\equiv\prod_{i\neq j}q_{ij}(w_{ij})
\end{equation}

\noindent by defining $p\equiv e^{-\theta}$. Note that this parameter has a precise probabilistic meaning, making even more evident the above prescription, $p<1$. In fact, $\langle a_{ij}\rangle=\sum_{w_{ij}=0}^{+\infty}a_{ij}q_{ij}(w_{ij})=p=1-q_{ij}(0)$. According to the maximum-likelihood principle \cite{mymethod,mylikelihood}, $p$ has to be calculated in terms of the observed quantities, by maximizing the function

\begin{equation}
\ln\mathcal{L}(\theta)=\ln P(\textbf{G}^*|\theta)=\sum_{i\neq j}\left[w_{ij}^*\ln(e^{-\theta})+\ln(1-e^{-\theta})\right]
\end{equation}

\noindent with respect to $\theta$. The solution to this optimization problem can be found by isolating $\theta$ in the above equation

\begin{equation}
W(\textbf{G}^*)=\sum_{i\neq j}\frac{e^{-\theta^*}}{1-e^{-\theta^*}}\equiv N(N-1)\frac{p^*}{1-p^*}=\langle W\rangle_{p^*}
\label{eq:likwrg}
\end{equation}

\noindent and, then, by inverting eq. \ref{eq:likwrg} (note that the condition expressed by eq. \ref{eq:con} is satisfied because $\langle w_{ij}\rangle=\sum_{w_{ij}=0}^{+\infty}w_{ij}q_{ij}(w_{ij})=\frac{p}{1-p}$):

\begin{equation}
p^*=\frac{W(\textbf{G}^*)}{W(\textbf{G}^*)+N(N-1)}.
\label{eq:p*}
\end{equation}

To calculate $\rho$ we need the expected value of $r$. Looking at its definition, we need the expected value of the minimum between $w_{ij}$ and $w_{ji}$:

\begin{equation}
\langle r\rangle\equiv\frac{\langle W^{\leftrightarrow}\rangle}{\langle W\rangle}=\frac{\sum_{i\neq j}\langle \min[w_{ij},\:w_{ji}]\rangle}{\langle W\rangle}.
\end{equation}

By considering that $w_{ij}$ and $w_{ji}$ are independent random variables, the cumulative distribution for the minimum is relatively easy to calculate:

\begin{equation}
P(\mbox{min}[w_{ij},\:w_{ji}]\geq w)=P(w_{ij}\geq w)P(w_{ji}\geq w)=p^{w}p^{w};
\end{equation}

\noindent from this, it follows that its expected value is

\begin{equation}
\langle \min[w_{ij},\:w_{ji}]\rangle_{WRG}=\sum^{+\infty}_{w=1}{P(\min[w_{ij},\:w_{ji}] \geq w)}={\frac{p^2}{1-p^2}}.
\label{eq:expminDWRG}
\end{equation}

Now, the expected value (that is, the ensemble average) of $r$, computed in correspondence of the maximum-likelihood parameters, can be found by using the result of eq. \ref{eq:p*}:

\begin{equation}
\langle r\rangle^*_{WRG}=\frac{\sum_{i\neq j}\frac{(p^*)^2}{1-(p^*)^2}}{\sum_{i\neq j} \frac{p^*}{1-p^*}}=\frac{p^*}{1+p^*}.
\end{equation}

\subsection{The Directed Weighted Configuration Model (WCM)}

This second null model is the weighted version of the Directed Configuration Model, fully specified by the in-degree and out-degree sequences \cite{newman_expo,mymethod}. The weighted counterparts of these constraints are the in-strength and out-strength sequences \cite{PRE2}:

\begin{equation}
H(\textbf{G}|\vec{\theta})=\sum_{i}(\alpha_{i} s_{i}^{out}+\beta_{i} s_{i}^{in})=\sum_{i\neq j}(\alpha_{i}+\beta_{j}) w_{ij};
\end{equation}

\noindent the partition function of the WCM is

\begin{eqnarray}
Z(\vec{\theta})&=&\sum_{\textbf{G}\in\mathcal{G}}e^{-H(\textbf{G}|\vec{\theta})}=\sum_{\textbf{G}\in\mathcal{G}}e^{\sum_{i\neq j}-(\alpha_{i}+\beta_{j}) w_{ij}}=\nonumber\\
&=&\prod_{i\neq j}\sum_{w_{ij}=0}^{+\infty}e^{-(\alpha_{i}+\beta_{j}) w_{ij}}=\prod_{i\neq j}\left[1-e^{-(\alpha_{i}+\beta_{j})}\right]^{-1}\nonumber
\end{eqnarray}

\noindent (where $\vec{\theta}\equiv\{\vec{\alpha},\:\vec{\beta}\}$ and provided that $e^{-(\alpha_{i}+\beta_{j})}<1$). Again, it is a product over $N(N-1)$ independent random variables. The reason becomes clearer when considering the WCM: when the contraints are \textit{local}, that is expressable as linear combinations of the adjacency matrix elements, the partition function factorizes and the probability of a given configuration factorizes as well, as a product of the independent random variables probability coefficients \cite{mymethod}. In this case every (non-negative, integer-valued) weighted network in the grandcanonical ensemble has a probability of the following form

\begin{eqnarray}
P(\textbf{G})&=&\frac{\prod_{i\neq j}e^{-(\alpha_{i}+\beta_{j}) w_{ij}}}{\prod_{i\neq j}\left[1-e^{-(\alpha_{i}+\beta_{j})}\right]^{-1}}\equiv\prod_{i\neq j}p_{ij}^{w_{ij}}(1-p_{ij})\equiv\nonumber\\
&\equiv&\prod_{i\neq j}(x_{i}y_{j})^{w_{ij}}(1-x_{i}y_{j})\equiv\prod_{i\neq j}q_{ij}(w_{ij})
\end{eqnarray}

\noindent by defining $p_{ij}\equiv e^{-(\alpha_{i}+\beta_{j})}=e^{-\alpha_i}e^{-\beta_j}\equiv x_{i}y_{j}$. Now, two parameters per vertex have to be calculated in terms of the observed quantities: the maximum-likelihood principle \cite{mymethod} prescribes to maximize

\begin{equation}
\ln\mathcal{L}(\vec{\theta})=\ln P(\textbf{G}^*|\vec{\theta})=\sum_{i\neq j}\left[w_{ij}^*\ln(x_{i}y_{j})+\ln(1-x_{i}y_{j})\right]
\end{equation}

\noindent with respect to $\vec{x}$ and $\vec{y}$. The solution to the optimization problem can be found by solving the system

\begin{eqnarray}
\left\{ \begin{array}{ll}
s^{out}_i(\textbf{G}^*) &= \sum_{j\ne i}\frac{x^*_iy^*_j}{1-x^*_iy^*_j}=\langle s_{i}^{out}\rangle_{\vec{\theta}^*},\quad \forall i\\
s^{in}_i(\textbf{G}^*) &= \sum_{j\ne i}\frac{x^*_jy^*_i}{1-x^*_jy^*_i}=\langle s_{i}^{in}\rangle_{\vec{\theta}^*},\quad\:\:\forall i
\end{array} \right.
\label{eq:wcm}
\end{eqnarray}

\noindent (again, the condition expressed by eq. \ref{eq:con} is satisfied because $\langle w_{ij}\rangle=\sum_{w_{ij}=0}^{+\infty}w_{ij}q_{ij}(w_{ij})=\frac{p_{ij}}{1-p_{ij}}\equiv\frac{x_{i}y_{j}}{1-x_{i}y_{j}}$). The expected value of the minimum between $w_{ij}$ and $w_{ji}$ can be easily found by generalizing eq. \ref{eq:expminDWRG}

\begin{equation}
\langle \min[w_{ij},\:w_{ji}]\rangle_{WCM}=\frac{p_{ij}p_{ji}}{1-p_{ij}p_{ji}}
\end{equation}

\noindent and the expected value of $r$, computed in correspondence of the maximum-likelihood parameters, can be found by using the results of eq. \ref{eq:wcm}:

\begin{equation}
\langle r\rangle_{WCM}^*=\frac{\sum_{i\neq j}\frac{p_{ij}^*p_{ji}^*}{1-p_{ij}^*p_{ji}^*}}{\sum_{i\neq j} \frac{p_{ij}^*}{1-p_{ij}^*}}.
\label{r_WCM}
\end{equation}

\subsection{The Balanced Configuration Model (BCM)}

In addition to the WCM, we further developed a version of it that is intended to model networks where the observed differences between $s_i^{out}$ and $s_{i}^{in}$ are interpreted as statistical fluctuations around a balanced condition, i.e. $\langle s_i^{out}\rangle=\langle s_{i}^{in}\rangle$. We can start from the WCM equations, to specify them in this particular case. The condition $s_{i}^{out}\simeq s_i^{in}$ implies that $x_i\simeq y_i\equiv z_{i}$ and this reduce the number of equations to solve, from $2N$ to $N$:

\begin{equation}
s_{i}^{out}+s_{i}^{in}=\sum_{j\neq i}\frac{2 z_{i}z_{j}}{1-z_{i}z_{j}}\Longrightarrow s_{i}^{tot}(\textbf{G}^*)=\sum_{j\ne i}\frac{2 z_{i}^*z_{j}^*}{1-z_{i}^*z_{j}^*},\quad\forall i.
\end{equation} 

This, in turn, implies that $p_{ij}=p_{ji}=z_{i}z_{j}$ and that $\langle w_{ij}\rangle=\langle w_{ji}\rangle$. So, under the BCM, the expected value of the minimum and of $r$ become, respectively,

\begin{equation}
\langle \min[w_{ij},\:w_{ji}]\rangle_{BCM}=\frac{p_{ij}^2}{1-p_{ij}^2}
\end{equation}

\noindent and

\begin{equation}
\langle r\rangle_{BCM}^*=\frac{\sum_{i<j}\frac{(p_{ij}^*)^{2}}{1-(p_{ij}^*)^{2}}}{\sum_{i<j}\frac{p_{ij}^*}{1-p_{ij}^*}}
\label{nullWBCM}
\end{equation}

\noindent which is nothing more that a simplified version of eq. \ref{r_WCM}.

The fundamental insight given by the BCM is that, in networks with node balance, i.e. where $\langle s_i^{out}\rangle=\langle s_{i}^{in}\rangle$, the expected weights are symmetric:

\begin{equation}
\langle w_{ij}\rangle_{BCM}^*=\frac{z_i z_j}{1-z_i z_j}=\langle w_{ji}\rangle_{BCM}^*
\end{equation}
This means that, in networks where the observed differences between $s_i^{out}$ and $s_{i}^{in}$ are consistent with statistical fluctuations around a balanced condition, one automatically expects symmetric weights, even without introducing any tendency towards reciprocation. 
This shows that measures of reciprocity based on the symmetry of mutual weights necessarily receive spurious contributions from other sources of flow balance.
This observation concludes our statement that previously attempted correlation- and symmetry-based measures \cite{jari,fagiolo,achen,faloutsos} cannot properly separate reciprocity from other factors.

\section{From null models to true models}
The three previous null models are defined in terms of constraints as the total weight, the in- and out-strength sequences and the total-strength sequence. So, not being included in the list of the constraints, the reciprocity $r$ and the index $\rho$ were a sort of target quantities, to test the power of the considered null models in reproducing them.

Now, we can make a step forward and include some information about the reciprocity structure of the network. 

\subsection{The Weighted Reciprocity Model (WRM)} 

We start by generalizing the WCM, by adding to its hamiltonian a sort of ``global reciprocity'' defined over the whole network, thus fixing the total number of reciprocal links. This means to consider, as a further constraint, the quantity

\begin{equation}
W^{\leftrightarrow}=\sum_{i\neq j}\min[w_{ij},\:w_{ji}]=\sum_{i}s_{i}^{\leftrightarrow}
\end{equation}

\noindent to obtain the following Hamiltonian

\begin{equation}
H(\mathbf{G}|\vec{\theta})=\sum_{i}(\alpha_{i}s_{i}^{out}+\beta_{i}s_{i}^{in})+\gamma W^{\leftrightarrow}
\end{equation}

\noindent (where $\vec{\theta}\equiv\{\vec{\alpha},\:\vec{\beta},\:\gamma\}$). The resolution of this null model is considerably simplified by considering an equivalent way of rewriting it,

\begin{equation}
H(\mathbf{G}|\vec{\theta})=\sum_{i}[\alpha_{i}s_{i}^{\rightarrow}+\beta_{i}s_{i}^{\leftarrow}+(\alpha_{i}+\beta_{i}+\gamma) s_{i}^{\leftrightarrow}]
\end{equation}

\noindent having posed, to uniform the formalism, $\min[w_{ij},\:w_{ji}]\equiv w_{ij}^{\leftrightarrow}$ and having defined

\begin{eqnarray}
s_{i}^{out}\equiv s_{i}^{\rightarrow}+s_{i}^{\leftrightarrow}&\Longrightarrow& w_{ij}\equiv w_{ij}^{\rightarrow}+w_{ij}^{\leftrightarrow},\nonumber\\
s_{i}^{in}\equiv s_{i}^{\leftarrow}+s_{i}^{\leftrightarrow}&\Longrightarrow& w_{ij}\equiv w_{ji}^{\leftarrow}+w_{ij}^{\leftrightarrow}.
\label{s_only}
\end{eqnarray}

\noindent Now, the most challenging calculation is about the partition function. This can be done by rewriting the hamiltonian solely in terms of the variables $w_{ij}^{\rightarrow}$, $w_{ij}^{\leftarrow}$ and $w_{ij}^{\leftrightarrow}$,

\begin{eqnarray}
H(\mathbf{G}|\vec{\theta})&=&\sum_{i<j}[(\alpha_{i}+\beta_{j})w_{ij}^{\rightarrow}+(\alpha_{j}+\beta_{i})w_{ij}^{\leftarrow}+\nonumber\\
&+&(\alpha_{i}+\beta_{j}+\alpha_{j}+\beta_{i}+2\gamma)w_{ij}^{\leftrightarrow}]
\end{eqnarray}

\noindent and considering the admissible states for them:

\begin{equation}
(w_{ij}^{\rightarrow},\:w_{ij}^{\leftarrow},\:w_{ij}^{\leftrightarrow})=\{(0,\:0,\:\mathbf{N}),\:(\mathbf{N}^+,\:0,\:\mathbf{N}),\:(0,\:\mathbf{N}^+,\:\mathbf{N})\}
\label{states}
\end{equation}

\noindent where $\mathbf{N}\equiv [0\dots\infty)$ and $\mathbf{N}^+\equiv [1\dots\infty)$. So the partition function becomes

\begin{eqnarray}
Z(\vec{\theta})&=&\sum_{\textbf{G}\in\mathcal{G}}e^{-H(\mathbf{G}|\vec{\theta})}=\sum_{(w_{ij}^{\rightarrow},\:w_{ij}^{\leftarrow},\:w_{ij}^{\leftrightarrow})}e^{-H(\mathbf{G}|\vec{\theta})}=\nonumber\\
&=&\prod_{i<j}\frac{(1-x_{i}x_{j}y_{i}y_{j})}{(1-x_{i}y_{j})(1-x_{j}y_{i})(1-x_{i}x_{j}y_{i}y_{j}z^2)}\equiv\nonumber\\
&\equiv&\prod_{i<j}Z_{ij}^{WRM}(\vec{\theta})
\end{eqnarray}

\noindent (having posed $x_{i}\equiv e^{-\alpha_{i}}$, $y_{i}\equiv e^{-\beta_{i}}$ and $z\equiv e^{-\gamma}$) and, consequently, the probability coefficient for the generic configuration $\mathbf{G}$ is

\begin{equation}
P(\mathbf{G})=\prod_{i<j}\frac{(x_{i}y_{j})^{w_{ij}}(x_{j}y_{i})^{w_{ji}}z^{2w_{ij}^{\leftrightarrow}}}{Z_{ij}^{WRM}(\vec{\theta})}.
\end{equation}

Now, the maximum-likelihood principle prescribes to maximize

\begin{eqnarray}
\ln P(\textbf{G}^*|\vec{\theta})&=&\sum_{i< j}[w_{ij}^*\ln(x_{i}y_{j})+w_{ji}^*\ln(x_{j}y_{i})+\nonumber\\
&+&(2w_{ij}^{\leftrightarrow})^*\ln z-\ln Z_{ij}^{WRM}(\vec{\theta})]
\end{eqnarray}

\noindent with respect to $\vec{x}$, $\vec{y}$ and $z$. The solution to the previous optimization problem can be found by solving the system

\begin{eqnarray}
\left\{ \begin{array}{ll}
s^{out}_i(\textbf{G}^*) &= \sum_{j\ne i}\langle w_{ij}\rangle_{\vec{\theta}^*}=\langle s_{i}^{out}\rangle_{\vec{\theta}^*},\quad \forall i\\
s^{in}_i(\textbf{G}^*) &=\sum_{j\ne i}\langle w_{ji}\rangle_{\vec{\theta}^*}=\langle s_{i}^{in}\rangle_{\vec{\theta}^*},\quad \:\:\forall i\\
W^{\leftrightarrow}(\textbf{G}^*) &= \sum_{i< j}2\langle w_{ij}^{\leftrightarrow}\rangle_{\vec{\theta}^*} = \langle W^{\leftrightarrow}\rangle_{\vec{\theta}^*}
\end{array} \right.
\label{eq:wrm}
\end{eqnarray}

\noindent where

\begin{equation}
\langle w_{ij}\rangle_{\vec{\theta}^*}=\frac{x^*_iy^*_j(1-x^*_jy^*_i)}{(1-x^*_iy^*_j)(1-x^*_ix^*_jy^*_iy^*_j)}+\frac{x^*_ix^*_jy^*_iy^*_j(z^*)^2}{1-x^*_ix^*_jy^*_iy^*_j(z^*)^2},
\end{equation}
\begin{equation}
\langle w_{ji}\rangle_{\vec{\theta}^*}=\frac{x^*_jy^*_i(1-x^*_iy^*_j)}{(1-x^*_jy^*_i)(1-x^*_ix^*_jy^*_iy^*_j)}+\frac{x^*_ix^*_jy^*_iy^*_j(z^*)^2}{1-x^*_ix^*_jy^*_iy^*_j(z^*)^2},
\end{equation}
\begin{equation}
\langle w_{ij}^{\leftrightarrow}\rangle_{\vec{\theta}^*}=\frac{x^*_ix^*_jy^*_iy^*_j(z^*)^2}{1-x^*_ix^*_jy^*_iy^*_j(z^*)^2}.
\end{equation}

Now, the expected value of the minimum between $w_{ij}$ and $w_{ji}$ is $\langle \min[w_{ij},\:w_{ji}]\rangle_{WRM}^*=\langle w_{ij}^{\leftrightarrow}\rangle_{\vec{\theta}^*}$. Even if it is possible to write down the analytical expression of the expected value of $r$, by using it, this can be avoided, by considering that

\begin{equation}
\langle r\rangle_{WRM}^*=\frac{\langle W^{\leftrightarrow}\rangle_{\vec{\theta}^*}}{\langle W\rangle_{\vec{\theta}^*}}=\frac{W^{\leftrightarrow}(\mathbf{G}^*)}{W(\mathbf{G}^*)}=r;
\end{equation}

\noindent this, in turn, implies that

\begin{equation}
\rho_{WRM}^*=\frac{r-\langle r\rangle_{WRM}^*}{1-\langle r\rangle_{WRM}^*}\equiv\frac{r-r}{1-r}=0.
\end{equation}

So, by definition, the index $\rho$ is trivially reproduced by the WRM. 

Note also that the only difference between the predicted quantities $\langle w_{ij}^{\leftrightarrow}\rangle_{WCM}$ and $\langle w_{ij}^{\leftrightarrow}\rangle_{WRM}$ lies in the presence of the extra-parameter $z$ in the second expression. Recalling that $z<1$, if the hidden variables $\vec{x}$ and $\vec{y}$ are kept fixed, changing $z$ means lowering the expected reciprocal weight with respect to the WCM prediction. This makes the WRM best suited to reproduce networks that are anti-reciprocal (i.e., less reciprocal than the WCM prediction).

\subsection{The Non-Reciprocated Strength Model (NSM)}

A second null model including the information about the global reciprocity structure of the network can be defined, starting by the WRM hamiltonian. This time, the imposed constraints are the in- and out-strength sequences, diminished by the reciprocal strength sequence (see eq. \ref{s_only}), and the total number of reciprocal links:

\begin{eqnarray}
H(\mathbf{G}|\vec{\theta})&=&\sum_{i}(\alpha_{i} s_{i}^{\rightarrow}+\beta_{i} s_{i}^{\leftarrow})+\gamma W^{\leftrightarrow}=\nonumber\\
&=&\sum_{i<j}[(\alpha_{i}+\beta_{j})w_{ij}^{\rightarrow}+(\alpha_{j}+\beta_{i})w_{ij}^{\leftarrow}+2\gamma w_{ij}^{\leftrightarrow}].\nonumber
\end{eqnarray}

Following the calculations of the WRM, the partition function is

\begin{eqnarray}
Z(\vec{\theta})&=&\sum_{\textbf{G}\in\mathcal{G}}e^{-H(\mathbf{G}|\vec{\theta})}=\sum_{(w_{ij}^{\rightarrow},\:w_{ij}^{\leftarrow},\:w_{ij}^{\leftrightarrow})}e^{-H(\mathbf{G}|\vec{\theta})}=\nonumber\\
&=&\prod_{i<j}\frac{(1-x_{i}x_{j}y_{i}y_{j})}{(1-x_{i}y_{j})(1-x_{j}y_{i})(1-z^2)}\equiv\nonumber\\
&\equiv&\prod_{i<j}Z_{ij}^{NSM}(\vec{\theta})
\end{eqnarray}

\noindent (having posed $x_{i}\equiv e^{-\alpha_{i}}$, $y_{i}\equiv e^{-\beta_{i}}$ and $z\equiv e^{-\gamma}$). The probability coefficient for a generic configuration, $\mathbf{G}$, is

\begin{equation}
P(\mathbf{G})=\prod_{i<j}\frac{(x_{i}y_{j})^{w_{ij}^{\rightarrow}}(x_{j}y_{i})^{w_{ij}^{\leftarrow}}z^{2w_{ij}^{\leftrightarrow}}}{Z_{ij}^{NSM}(\vec{\theta})}
\end{equation}

\noindent and the maximum-likelihood principle prescribes to maximize

\begin{eqnarray}
\ln P(\textbf{G}^*|\vec{\theta})&=&\sum_{i< j}[(w_{ij}^{\rightarrow})^*\ln(x_{i}y_{j})+(w_{ij}^{\leftarrow})^*\ln(x_{j}y_{i})+\nonumber\\
&+&(2w_{ij}^{\leftrightarrow})^*\ln z-\ln Z_{ij}^{NSM}(\vec{\theta})]
\end{eqnarray}

\noindent with respect to $\vec{x}$, $\vec{y}$ and $z$. The solution to the previous optimization problem can be found by solving the system

\begin{eqnarray}
\left\{ \begin{array}{ll}
s^{\rightarrow}_i(\textbf{G}^*) &= \sum_{j\ne i}\langle w_{ij}^{\rightarrow}\rangle_{\vec{\theta}^*}=\langle s_{i}^{\rightarrow}\rangle_{\vec{\theta}^*},\quad \forall i\\
s^{\leftarrow}_i(\textbf{G}^*) &=\sum_{j\ne i}\langle w_{ij}^{\leftarrow}\rangle_{\vec{\theta}^*}=\langle s_{i}^{\leftarrow}\rangle_{\vec{\theta}^*},\quad \forall i\\
W^{\leftrightarrow}(\textbf{G}^*) &= \sum_{i< j}2\langle w_{ij}^{\leftrightarrow}\rangle_{\vec{\theta}^*} = \langle W^{\leftrightarrow}\rangle_{\vec{\theta}^*}
\end{array} \right.
\label{eq:wrm}
\end{eqnarray}

\noindent where

\begin{equation}
\langle w_{ij}^{\rightarrow}\rangle_{\vec{\theta}^*}=\frac{x^*_iy^*_j(1-x^*_jy^*_i)}{(1-x^*_iy^*_j)(1-x^*_ix^*_jy^*_iy^*_j)}+\frac{(z^*)^2}{1-(z^*)^2},
\end{equation}
\begin{equation}
\langle w_{ij}^{\leftarrow}\rangle_{\vec{\theta}^*}=\frac{x^*_jy^*_i(1-x^*_iy^*_j)}{(1-x^*_jy^*_i)(1-x^*_ix^*_jy^*_iy^*_j)}+\frac{(z^*)^2}{1-(z^*)^2},
\end{equation}
\begin{equation}
\langle w_{ij}^{\leftrightarrow}\rangle_{\vec{\theta}^*}=\frac{(z^*)^2}{1-(z^*)^2}.
\end{equation}

As for the WRM

\begin{equation}
\langle r\rangle_{NSM}^*=\frac{\langle W^{\leftrightarrow}\rangle_{\vec{\theta}^*}}{\langle W\rangle_{\vec{\theta}^*}}=\frac{W^{\leftrightarrow}(\mathbf{G}^*)}{W(\mathbf{G}^*)}=r;
\end{equation}

\noindent this, in turn, implies that

\begin{equation}
\rho_{NSM}^*=\frac{r-\langle r\rangle_{NSM}^*}{1-\langle r\rangle_{NSM}^*}\equiv\frac{r-r}{1-r}=0.
\end{equation}

\subsection{The Reciprocated Strength Model (RSM)} 

Until now, we have defined three null models with no constraints about the reciprocity (the WRG, the WCM and the BCM) and two null models with the total number of reciprocal links (thus implementing a global notion of reciprocity), as a constraint.

Now, we can define more refined null models, by considering, as constraints, the local notion of reciprocity, as defined by eq. \ref{srec}. We start by considering the following hamiltonian:

\begin{equation}
H(\mathbf{G}|\vec{\theta})=\alpha W+\sum_{i}\delta_{i}s_{i}^{\leftrightarrow}
\end{equation}

\noindent (where $\vec{\theta}\equiv\{\alpha,\:\vec{\delta}\}$). The resolution of the null model described by this hamiltonian is, again, considerably simplified by considering the equivalent expression

\begin{equation}
H(\mathbf{G}|\vec{\theta})=\sum_{i<j}[\alpha w_{ij}^{\rightarrow}+\alpha w_{ij}^{\leftarrow}+(2\alpha+\delta_{i}+\delta_{j})w_{ij}^{\leftrightarrow}];
\end{equation}

\noindent by summing over the states defined in eq. \ref{states} we find the partition function

\begin{eqnarray}
Z(\vec{\theta})&=&\sum_{\textbf{G}\in\mathcal{G}}e^{-H(\mathbf{G}|\vec{\theta})}=\sum_{(w_{ij}^{\rightarrow},\:w_{ij}^{\leftarrow},\:w_{ij}^{\leftrightarrow})}e^{-H(\mathbf{G}|\vec{\theta})}=\nonumber\\
&=&\prod_{i<j}\frac{(1+x)}{(1-x)(1-x^2 z_{i}z_{j})}
\end{eqnarray}

\noindent (having posed $x\equiv e^{-\alpha}$, $z_{i}\equiv e^{-\delta_{i}}$) the probability coefficient for the generic configuration $\mathbf{G}$ is

\begin{eqnarray}
P(\mathbf{G})&=&\prod_{i<j}\frac{x^{w_{ij}^{\rightarrow}+w_{ij}^{\leftarrow}+2w_{ij}^{\leftrightarrow}}(z_{i}z_{j})^{w_{ij}^{\leftrightarrow}}(1-x^2z_{i}z_{j})(1-x)}{(1+x)}=\nonumber\\
&=&\prod_{i<j}\frac{x^{w_{ij}+w_{ji}}(z_{i}z_{j})^{w_{ij}^{\leftrightarrow}}(1-x^2z_{i}z_{j})(1-x)}{(1+x)}
\end{eqnarray}

\noindent and the likelihood function is, of course, the logarithm of the previous probability coefficient. The solution to this optimization problem prescribes to solve the following system

\begin{eqnarray}
\left\{ \begin{array}{ll}
s^{\leftrightarrow}_i(\textbf{G}^*) &=\sum_{j\ne i}\langle w_{ij}^{\leftrightarrow}\rangle_{\vec{\theta}^*}=\langle s^{\leftrightarrow}_i\rangle_{\vec{\theta}^*},\quad \:\:\forall i\\
W(\textbf{G}^*) &= \sum_{i\neq j}\langle w_{ij}\rangle_{\vec{\theta}^*}=\langle W\rangle_{\vec{\theta}^*}
\end{array} \right.
\label{eq:rcm}
\end{eqnarray}

\noindent where

\begin{equation}
\langle w_{ij}\rangle_{\vec{\theta}^*}=\frac{x^*}{(1-(x^*)^2)}+\frac{(x^*)^2z^*_iz^*_j}{1-(x^*)^2z^*_iz^*_j},
\label{wDWRM}
\end{equation}
\begin{equation}
\langle w_{ij}^{\leftrightarrow}\rangle_{\vec{\theta}^*}=\frac{(x^*)^2z^*_iz^*_j}{1-(x^*)^2z^*_iz^*_j}.
\end{equation}

This model allows to solve for the $x$ value analitically. In fact, by summing eq. \ref{wDWRM} over the ordered pairs of nodes, we find

\begin{equation}
W(\mathbf{G}^*)=\frac{N(N-1) x^*}{(1-(x^*)^2)}+W^{\leftrightarrow}(\mathbf{G}^*)
\end{equation}

\noindent and by solving this second-order equation w.r.t. $x$, and taking the positive solution, we have the maximum-likelihood estimation of this parameter. Also this model exactly reproduces the observed reciprocity, because $W^{\leftrightarrow}(\mathbf{G}^*)=\langle W^{\leftrightarrow}\rangle_{\vec{\theta}^*}$ and  $W(\mathbf{G}^*)=\langle W\rangle_{\vec{\theta}^*}$. This means that $\rho_{RSM}=\frac{r-\langle r\rangle_{RSM}}{1-\langle r\rangle_{RSM}}\equiv\frac{r-r}{1-r}=0$ and the local quantities as the reciprocal strength sequence are now trivially reproduced.

\subsection{The Weighted Reciprocated Configuration Model (WRCM)}

The last step is the definion of a very general null model, to finally include those local quantities not fixed by the NSM and the RSM. This implies a slight generalization of the formulas in the previous two paragraphs. The graph hamiltonian becomes

\begin{equation}
H(\mathbf{G}|\vec{\theta})=\sum_{i}(\alpha_{i}s_{i}^{\rightarrow}+\beta_{i}s_{i}^{\leftarrow}+\gamma_{i} s_{i}^{\leftrightarrow})
\end{equation}

\noindent where, now, $\vec{\theta}\equiv\{\vec{\alpha},\:\vec{\beta},\:\vec{\gamma}\}$ and

\begin{eqnarray}
s_{i}^{\rightarrow}\equiv \sum_{j(\neq i)}w_{ij}^{\rightarrow},\:s_{i}^{\leftarrow}\equiv \sum_{j(\neq i)}w_{ij}^{\leftarrow},\:s_{i}^{\leftrightarrow}\equiv \sum_{j(\neq i)}w_{ij}^{\leftrightarrow}
\end{eqnarray}

\noindent with obvious meaning of the symbols (defined above). The partition function now becomes

\begin{eqnarray}
Z(\vec{\theta})&=&\prod_{i<j}\frac{(1-x_{i}x_{j}y_{i}y_{j})}{(1-x_{i}y_{j})(1-x_{j}y_{i})(1-z_{i}z_{j})}\equiv\nonumber\\
&\equiv&\prod_{i<j}Z_{ij}^{WRCM}(\vec{\theta})
\end{eqnarray}

\noindent and the likelihood is

\begin{eqnarray}
\ln P(\textbf{G}^*|\vec{\theta})&=&\sum_{i< j}[(w_{ij}^{\rightarrow})^*\ln(x_{i}y_{j})+(w_{ij}^{\leftarrow})^*\ln(x_{j}y_{i})+\nonumber\\
&+&(w_{ij}^{\leftrightarrow})^* \ln(z_{i}z_{j})-\ln Z_{ij}^{WRCM}(\vec{\theta})].
\end{eqnarray}

The solution to this optimization problem, with respect to $\vec{x}$, $\vec{y}$ and $\vec{z}$, can be found by solving the following system:

\begin{eqnarray}
\left\{ \begin{array}{ll}
s^{\rightarrow}_i(\textbf{G}^*) &= \sum_{j\ne i}\langle w_{ij}^{\rightarrow}\rangle_{\vec{\theta}^*}=\langle s_{i}^{\rightarrow}\rangle_{\vec{\theta}^*},\quad \forall i\\
s^{\leftarrow}_i(\textbf{G}^*) &=\sum_{j\ne i}\langle w_{ij}^{\leftarrow}\rangle_{\vec{\theta}^*}=\langle s_{i}^{\leftarrow}\rangle_{\vec{\theta}^*},\quad \forall i\\
s_{i}^{\leftrightarrow}(\textbf{G}^*) &= \sum_{j\ne i}\langle w_{ij}^{\leftrightarrow}\rangle_{\vec{\theta}^*} = \langle s_{i}^{\leftrightarrow}\rangle_{\vec{\theta}^*},\quad \forall i
\end{array} \right.
\label{eq:wrcm}
\end{eqnarray}

\noindent where

\begin{equation}
\langle w_{ij}^{\rightarrow}\rangle_{\vec{\theta}^*}=\frac{x^*_iy^*_j(1-x^*_jy^*_i)}{(1-x^*_iy^*_j)(1-x^*_ix^*_jy^*_iy^*_j)},
\end{equation}
\begin{equation}
\langle w_{ij}^{\leftarrow}\rangle_{\vec{\theta}^*}=\frac{x^*_jy^*_i(1-x^*_iy^*_j)}{(1-x^*_jy^*_i)(1-x^*_ix^*_jy^*_iy^*_j)},
\end{equation}
\begin{equation}
\langle w_{ij}^{\leftrightarrow}\rangle_{\vec{\theta}^*}=\frac{z^*_iz^*_j}{1-z^*_iz^*_j}.
\end{equation}

By the definition of the WRCM model, we not only recover the result that the global reciprocity is equal to the observed one (implying $r\equiv \langle r\rangle_{WRCM}$ and $\rho_{WRCM}\equiv0$, also valid for the WRM): now, all the vertex-level, strength sequences are exactly reproduced, impyling that the reciprocity is reproduced at a \emph{local} level.

The WRCM is now powerful enough to allow for the analysis of the weighted motifs (to understand which all the dyadic information has to be fixed) and for the community detection, especially for those networks where the reciprocity plays an important role in shaping its structure.

\section{Models: a summary}

The first six models explained in the previous sections can be recovered from the last and most general one (the WRCM) by means of simple substitutions in the graph hamiltonian (as shown by the following table).

\begin{center}
\begin{tabular}{|l|l|}
\hline
\multicolumn{2}{|c|}{\textit{Non-reciprocal models}}\\
\hline
$\alpha_{i}=\beta_{i}=\gamma;\:\gamma_{i}=2\gamma,\:\forall i$ & $\mbox{WRCM}\rightarrow\mbox{WRG}$\\
\hline
$\alpha_{i}=\beta_{i};\:\gamma_{i}=\alpha_{i}+\beta_{i},\:\forall i$ & $\mbox{WRCM}\rightarrow\mbox{BCM}$\\
\hline
$\alpha_{i}\neq \beta_{i};\:\gamma_{i}=\alpha_{i}+\beta_{i},\:\forall i$ & $\mbox{WRCM}\rightarrow\mbox{WCM}$\\
\hline
\multicolumn{2}{|c|}{\textit{Global-reciprocity models}}\\
\hline
$\alpha_{i}\neq \beta_{i};\:\gamma_{i}=\alpha_{i}+\beta_{i}+\gamma,\:\forall i$ & $\mbox{WRCM}\rightarrow\mbox{WRM}$\\
\hline
$\alpha_{i}\neq \beta_{i};\:\gamma_{i}=\gamma,\:\forall i$ & $\mbox{WRCM}\rightarrow\mbox{NSM}$\\
\hline
\multicolumn{2}{|c|}{\textit{Local-reciprocity models}}\\
\hline
$\alpha_{i}=\beta_{i}=\alpha;\:\gamma_{i}=\alpha_{i}+\beta_{i}+\delta_{i},\:\forall i$ & $\mbox{WRCM}\rightarrow\mbox{RSM}$\\
\hline
\end{tabular}
\end{center}

\section{The jackknife method}

The jackknife method \cite{jack,newman-mixing} is an expedient to mimic resampling and it is usually used to estimate the variance of a given function of the population mean, $f(\langle x\rangle)$ (being $x$ the random variable of interest). Doing this in the least biased way, would imply to have a whole collection of samples. However, we observe only a single realization. How can compensate for the lack of such observations? We can build a set of artificial samples by considering the following sets:

\begin{eqnarray}
s_{1}^J&=&\{x_{2},\:x_{3}\dots x_{M}\},\nonumber\\
s_{2}^J&=&\{x_{1},\:x_{3}\dots x_{M}\},\nonumber\\
&\vdots&\nonumber\\
s_{M}^J&=&\{x_{1},\:x_{2}\dots x_{M-1}\};
\end{eqnarray}

\noindent that is a list of vectors for each of which a single observation has been removed. Then, we calculate the so called \textit{jackknife averages}

\begin{equation}
\bar{s_{1}}^J=\frac{\sum_{i\neq 1}x_{i}}{M-1},\:\bar{s_{2}}^J=\frac{\sum_{i\neq 2}x_{i}}{M-1}\dots \bar{s_{M}}^J=\frac{\sum_{i\neq M}x_{i}}{M-1},
\end{equation}

\noindent the estimates of the first two moments

\begin{equation}
\mu_{1}^J\equiv\frac{\sum_{i}f(\bar{s_{i}}^J)}{M};\:\mu_{2}^J\equiv\frac{\sum_{i}f(\bar{s_{i}}^J)^2}{M},
\end{equation}

\noindent from which the estimate of the jackkinfe-standard deviation follows

\begin{equation}
\sigma_{f}^J\simeq \sqrt{\mu_{2}^J-(\mu_{1}^J)^2},
\end{equation}

\noindent and, finally \cite{newman-mixing},

\begin{equation}
\sigma_{f(\langle x\rangle)}\simeq \sqrt{M-1}\:\sigma_{f}^J.
\end{equation}

How can we implement all this for our weighted networks? The quantity we are interested in is $\rho$. It is a function of the expected value of $r$, taken over the whole grandcanonical ensemble: $\langle r\rangle$. By applying the jackknife method, we can build $L$ artificial samples by removing one \textit{weight} at a time. By rewriting the above formulas, the final estimates become

\begin{equation}
\rho_{NM}=\frac{r-\langle r\rangle_{NM}}{1-\langle r\rangle_{NM}},
\end{equation}
\begin{equation}
\sigma_{\rho_{NM}}^2=\sum_{i}^{L}(\rho_{i,\:NM}-\rho_{NM})^2=\frac{\sigma_{r}^2}{(1-\langle r_{NM}\rangle)^2},
\end{equation}

\noindent where $NM$ can be $WRG$, $WCM$, $BCM$, $WRM$, $NSM$, $RSM$, $WRCM$ and where the sum over the index $i$ means that we are summing over the realizations with the $i$-th weight removed.

\section{Description of the dataset}

In what follows a brief description of the analysed networks is given.

{\bf Interbank network.} This is the network of the Italian interbank monetary exchanges \cite{interbank}, in the year 1999. We analysed the monthy transactions for May ($N=215$, $L=5269$), June ($N=215$, $L=5229$), August ($N=215$, $L=5071$), October ($N=215$, $L=4712$) and December ($N=215$, $L=4685$). {\bf Food webs.} We analysed eight different food webs \cite{pajek2,pajek,foodwebmotifs}, from different ecosystems (lagoons, marshes, lakes, bays, estuaries, grasses, rivers), with a prevalence of aquatic habitats: {\it Chesapeake Bay} ($N=34$, $L=177$) and {\it Mondego Bay} ($N=46$, $L=400$), {\it Everglades Marshes} ($N=69$, $L=916$), {\it Maspalomas Lagoon} ($N=24$, $L=82$), {\it Michigan Lake} ($N=39$, $L=221$), {\it St. Marks Seagrass} ($N=54$, $L=536$), {\it Crystal River Creek} ($N=24$, $L=125$ and $N=24$, $L=100$). {\bf Neural networks.} We analysed the neural network \cite{neural} of \emph{C. Elegans} ($N=297$, $L=2345$). {\bf Social networks.} We analysed three different social networks \cite{pajek2,Social1,Social2,Social3,Social4,Social5,Social6}: {\it BK-Office}, {\it BK-Tech} and {\it BK-Fraternity}. BK-Tech and BK-Fraternity are completely connected (that is, $L=N(N-1)$). Bernard and Killworth (and, later, also with the help of Sailer), collected five sets of data on human interactions in bounded groups. {\it BK-Office} ($N=40$, $L=1558$) is the network of the human interactions (conversations) frequency between the employees of a small business-office, as recorded at time intervals of fifteen minutes (during two four-days periods), by an external observer, along a fixed route through the office itself. {\it BK-Tech} ($N=34$, $L=1122$) is the network of the human interactions (conversations) frequency between collaborators in a technical research group at a West Virginia University, as recorded at time intervals of half-hour (during one five-days working week), by an external observer. {\it BK-Fraternity} ($N=58$, $L=3306$) is the network of the human interactions (conversations) frequency between the students living in a fraternity at a West Virginia College, as recorded by an external observer at time intervals of fifteen minutes (during a five-days week, twenty-one hours per day) who walked through the public areas of the building. {\bf The World Trade Network.} We analyse the series of yearly bilateral data on exports and imports among world countries from the database in ref.\cite{wtw}, from 1948 to 2000 ($N\in[82,\:186]$ and $L\in[2539,\:19903]$).

\end{document}